\documentclass[iop]{emulateapj}

\newcommand{\bvec}[1]{\textbf{#1}}

\newcommand{\mytilde}{\raise.19ex\hbox{$\scriptstyle\sim$}}

\shorttitle{HST Study of Dark Core in Abell 520}
\shortauthors{Jee et al.}

\begin{document}

\title{  \textbf{\emph{HUBBLE SPACE TELESCOPE}}/ADVANCED CAMERA FOR SURVEYS 
CONFIRMATION OF THE DARK SUBSTRUCTURE IN A520\altaffilmark{1}}

\altaffiltext{1}{ Based on observations made with the NASA/ESA {\it  Hubble Space Telescope}, obtained at the Space Telescope Science Institute, which is operated by the Association of Universities for Research in Astronomy, Inc.} 

\author{M.~J.~JEE\altaffilmark{2}, H..~HOEKSTRA\altaffilmark{3}, A. MAHDAVI\altaffilmark{4}, AND  A.~BABUL\altaffilmark{5,6}}

\altaffiltext{2}{Department of Physics, University of California, Davis, One Shields Avenue, Davis, CA 95616, USA}
\altaffiltext{3}{Leiden Observatory, Leiden University, Leiden, The Netherlands}
\altaffiltext{4}{Department of Physics and Astronomy, San Francisco State University, San Francisco, CA 94131, USA}
\altaffiltext{5}{Department of Physics and Astronomy, University of Victoria, Victoria, BC, Canada}
\altaffiltext{6}{Kavli Institute for Theoretical Physics, Kohn Hall, University of California, Santa Barbara, CA 93106, USA}

\begin{abstract}

We present the results from a weak gravitational lensing study of the
merging cluster A520 based on the analysis of {\it Hubble Space
  Telescope}/Advanced Camera for Surveys (ACS) data. The excellent
data quality allows us to reach a mean number density of source
galaxies of $\mytilde109$ per sq. arcmin, which improves both
resolution and significance of the mass reconstruction compared to a
previous study based on Wide Field Planetary Camera 2 (WFPC2) images.
We take care in removing instrumental effects such as the
trailing of charge due to radiation damage of the ACS detector and the
position-dependent point spread function (PSF).
This new ACS analysis confirms the previous claims that a substantial
amount of dark mass is present between two luminous subclusters. We
examine the distribution of cluster galaxies and observe very little
light at this location. We find that the centroid of the dark peak in
the current ACS analysis is offset to the southwest by $\mytilde1\arcmin$
with respect to the centroid from the WFPC2 analysis. Interestingly,
this new centroid is in better spatial agreement with the location
where the X-ray emission is strongest, and the mass-to-light ratio
estimated with this centroid is much higher ($813\pm78
M_{\sun}/L_{R\sun}$) than the previous value; the aperture mass based
on the WFPC2 centroid provides a slightly lower, but consistent
mass. Although we cannot provide a definite explanation for the
presence of the dark peak, we discuss a revised scenario, wherein dark
matter with a more conventional range ($\sigma_{DM}/m_{DM} < 1 \mbox{
  cm}^{2} \mbox{g}^{-1}$) of self-interacting cross-section can lead
to the detection of this dark substructure. If supported by detailed
numerical simulations, this hypothesis opens up the possibility that
the A520 system can be used to establish a lower limit of the
self-interacting cross-section of dark matter.

\end{abstract}

\keywords{
gravitational lensing ---
dark matter ---
cosmology: observations ---
X-rays: galaxies: clusters ---
galaxies: clusters: individual (Abell 520) ---
galaxies: high-redshift}

\section{INTRODUCTION} \label{section_introduction}

Galaxy clusters are comprised of dark matter, cluster galaxies, and
hot plasma.  When two clusters collide, it is believed that the
galaxies and dark matter temporarily dissociate from the hot plasma
because the latter is collisional and subject to ram
pressure. Eventually, the dark matter of the cluster, the
gravitationally dominant component, pulls the hot plasma back into its
potential well, and the dissociation disappears. Therefore, in general
we have a narrow time-window (a few Gyrs after the core pass-through)
to witness observationally significant offsets between collisional and
collisionless constituents of the clusters.

Detailed studies of these ``dissociative" mergers (Dawson 2012)
provide unique opportunities to enhance not only our astrophysical
understanding of the cluster formation and evolution, but also our
understanding of fundamental physics on the nature of dark
matter. Because a direct lab detection of dark matter particles may
not happen within the current decade, these merging clusters
with large offsets among the different cluster constituents are
receiving growing attention (e.g., Springel \& Farrar 2007; Randall et
al. 2008; Dawson 2012).

To date, only a few merging systems are known to possess such large
dissociative features (e.g., Dawson et al. 2012, Merten et al. 2011,
Bradac et al. 2008; Okabe \& Umetsu 2008; Soucail 2012), and only two
systems, namely 1E0657-56 at $z=0.3$ (Markevitch et al. 2002;
hereafter the ``Bullet Cluster") and A520 at $z=0.2$ (Markevitch et
al. 2005), are known to possess prominent X-ray ``bow-shock''
features, which serve as the definite evidence for a recent high-speed
collision and allow us to estimate the relatively stable transverse
velocity of the collision.  Clowe et al. (2006) showed that the mass
distribution of the Bullet Cluster revealed by weak-lensing closely
follows the cluster galaxies, which themselves are offset from the
X-ray emitting gas. The observation is claimed as proof of dark matter, and
some studies (e.g., Markevitch et al. 2004; Randall et al. 2008) use
the result to constrain an upper limit of the dark matter self-interaction
cross-section.

Mahdavi et al. (2007; hereafter M07) reported a more puzzling case
based on their weak-lensing analysis of A520 with Canada France
Hawaii Telescope (CFHT) and Subaru images. The mass distribution of
M07 generally follows the cluster galaxies as seen in the Bullet
Cluster. However, what separates A520 from the Bullet Cluster is the
presence of a ``dark core'', which coincides with the location of
the peak of the X-ray emission. The dark core region does not contain
any luminous cluster galaxies unlike the other mass peaks in M07.
Although the initial independent analysis of A520 with Subaru images
by Okabe \& Umetsu (2008) did not support the presence of the dark
core, their final study carried out with the subset of their data
(taken with the auto guider of the telescope turned on) agrees with
M07 in that their weak lensing data shows the dark peak near the peak
of the X-ray emission.  This study of Okabe \& Umetsu (2008) serves as
a good example to illustrate that the investigation of cluster
substructures is prone to instrumental systematics such as imperfect
point spread function models.

A logical extension of these studies is a follow-up investigation with
space-based images, which provide a much higher density of source
galaxies. If the dark core had appeared as an ``unfortunate
statistical fluke'' due to fortuitous alignment of source galaxies,
mass reconstruction with a much higher source density must reduce
the probability of this chance alignment and thus the significance of
the feature.  In Jee et al. (2012; hereafter J12), we presented a {\it
  Hubble Space Telescope (HST)} Wide Field Planetary Camera 2 (WFPC2)
weak lensing analysis of A520 and confirmed the results of M07 and
Okabe \& Umetsu (2008). Both the two-dimensional mass map and the
aperture mass results of J12 are consistent with M07, except for the
detection of two additional mass peaks, which were not reported in
M07. One of the two new peaks (labeled as P5 in J12) resolves one of
the questions raised in M07, who considered the absence of any
significant mass peak around this location also discordant with our
common light-traces-mass hypothesis along with the presence of the
significant mass in the dark core region.  The other new peak (labeled
as P6 in J12) is detected $\mytilde200$~kpc south of the dark core and
coincides with the spatial distribution of the cluster galaxies.  As
for the dark core both J12 and M07 discuss a number of possibilities
that may lead to the observation, but neither study could exclusively
single out one definite scenario responsible for these observations.

The scenarios considered in J12 and M07 are: 1) a possible
presence of a background high-redshift ($z>1$) cluster at the
location of the dark peak, 2) a rejection of cluster galaxies during a complex
multi-body collision, 3)  a filament elongated along the line-of-sight direction,
4) self-interacting dark matter, and 5) a compact high M/L group.
We reject the first scenario because our spectroscopic survey data
do not indicate the presence of a possible $z>1$ cluster at the dark peak, 
the X-ray emission line is consistent with the A520 redshift, and the dark peak
is seen in ground-based weak-lensing, where the mean source redshift
is lower than $z\sim1$.
Scenario 2 is discarded because no numerical simulations have shown that
brightest cluster galaxies are ejected during three-body encounter, although it
can happen to faint satellite galaxies. The third possibility is unlikely because
the filament must be very thin in such a way that most of the projected mass is confined to the
central $r<150$kpc region around the dark peak. 
The fourth scenario is highly disfavored because the anticipated collisional cross-section
of dark matter by M07 is much higher than the values estimated from the Bullet Cluster. 
The last scenario has not been completely ruled out yet. However, the implied M/L value of the group is
too high to be bracketed by the values found in the literature.

The improved sampling resolution and sensitivity of the Advanced Camera for
Surveys (ACS) allow for a more efficient coverage of A520 reaching a
higher source density in principle. Recently, Clowe et al. (2012;
hereafter C12) presented the results from their weak lensing analysis
of A520, this time using such ACS data. They reported that no such
dark peak is seen in their analysis. The rest of the weak lensing
substructures of C12 closely resemble those of J12 including the
aforementioned two new mass peaks of J12.  Although this claim of C12,
if true, may resolve the ``mystery'' reported by J12, M07, and Okabe
\& Umetsu (2008), the discrepancy raises a different kind of puzzle,
namely why is the result of the C12 weak lensing study so different
from those of three previous independent studies.
 
To address this issue, we perform an independent weak lensing analysis
of A520 using the same ACS data used by C12.  We provide a detailed
comparison of the results with those from C12.
The structure of the paper is as follows.
In \S2 we present the data and our analysis. Correction of
instrumental effects is detailed in \S3. The mass reconstruction is
discussed in \S4 and mass estimates are presented in \S5. We compare
our results to those of C12 in \S6 and examine whether the new observations
can constrain the interaction cross-section of dark matter in \S6.
Throughout this paper we use $(\Omega_M, \Omega_{\Lambda}, h) = (0.3,
0.7, 0.7)$ for cosmology unless explicitly stated otherwise. This
gives a plate scale of $\mytilde3.3$~kpc/$\arcsec$ at the redshift
($z=0.2$) of Abell 520. All the quoted uncertainties are at the
1-$\sigma$ ($\sim68$\%) level.

\section{DATA} \label{section_obs}

We retrieved the {\it HST}/ACS images of A520 (PI: Clowe)
from the Mikulski Archive for Space Telescopes
(MAST)\footnote{http://archive.stsci.edu} in
 2012 July after the new charge transfer 
inefficiency (CTI)\footnote{CTI
  causes systematic elongation of object shapes along the readout
  direction. Readers are referred to \ref{section_cti_impacts} for
  details.} 
correction of Ubeda and Anderson (2012) had become available.
The cluster was observed in the Cycle 18 (2011 February and April)
and the raw data are
severely affected by  CTI.
The images are comprised
of four pointings in F435W, F606W, and F814W to cover the
approximately $7\arcmin\times7\arcmin$ central region of the
cluster. The exposure time per pointing is 4,600 s for F814W whereas
it is a factor of two smaller for both F435W and F606W. A similar
exposure time per pointing (4,400 s) was used for the WFPC2
observation (PI: Dalcanton) of A520, although the ACS image is in
general $\mytilde$1.4 magnitude deeper because of its higher (more
than a factor of 3) sensitivity. However, around the dark core the
WFPC2 image provides comparable depth because the WFPC2 observation
was designed in such a way that the footprints overlap substantially
in that region, although the ACS weak lensing analysis benefits from the
improved sampling. We use the F814W data as the
primary data set for the lensing analysis. However, we verify that
a parallel weak-lensing analysis with the F606W image gives a consistent result 
yet with slightly increased noise because of the shallower depth.

We combine the four ACS pointings into a single mosaic image, which is
used to measure the galaxy shapes (note that we keep track of the
combined PSF). As discussed in more detail in \S2, a careful treatment
of the impact of the CTI is critical. To investigate this, we created three
different versions of mosaic images.  The first mosaic was generated
with the {\tt FLT} files processed by the CALACS pipeline (Hack et
al. 2003). These images are corrected for the bias stripping
noise\footnote{http://www.stsci.edu/hst/acs/software/destripe/} with
no CTI correction applied. The second mosaic is made from the {\tt
  FLC} files also processed by the CALACS pipeline, which removes both
bias stripping noise and CTI trails using the latest 2012 CTI model of
Ubeda and Anderson (2012).  This latest model was not available to
C12.  Finally, for our third mosaic image, we use the {\tt PixCteCorr}
script to correct the CTI effects as per the 2009 CTI model (Anderson
\& Bedin 2010).  This old model is considered inferior to the latest
model for the accuracy of the correction in the low-flux regime. We
defer the detailed comparison to
\textsection\ref{section_cti_impacts}.

The image offsets are computed by comparing the coordinates of common
astronomical sources. This process can be easily automated for the
image set observed at the same pointing with an accuracy of
$\lesssim0.01$ pixels. However, the offsets between different
pointings are difficult to determine reliably using automated
algorithms because the small overlapping areas ($\mytilde200$ pixels)
and the dense distribution of cosmic rays make only a few objects
available for shift estimation. Therefore, we choose to create a
separate stack for each pointing as an intermediate step and to use the catalog from the
resulting image (where cosmic rays are removed and fainter sources are
available) to determine accurate offsets between different pointings
($\mytilde0.02$ pixel).  The final full $2\times2$ mosaic is generated by the {\tt
  MultiDrizzle} software (Koekemoer et al. 2002) with the {\tt
  Lanczos3} drizzling kernel and an output pixel scale of $0\farcs05$.
The {\tt Lanczos3} kernel closely approximates the theoretically ideal
sinc ($\sin x/x$) interpolation kernel by truncating the oscillation
beyond the third pixel from the center.  In Jee et al. (2007a), we
demonstrate that this {\tt Lanczos3} drizzling kernel is superior to
the  ``square'' kernels in minimizing both aliasing and noise correlation 
and gives the sharpest PSF. We note that C12 used a square kernel with
an output pixel scale of $0\farcs05$ to drizzle the ACS images, which
may not be optimal for measuring accurate shapes of small galaxies.

\begin{figure*}
\includegraphics[width=6cm]{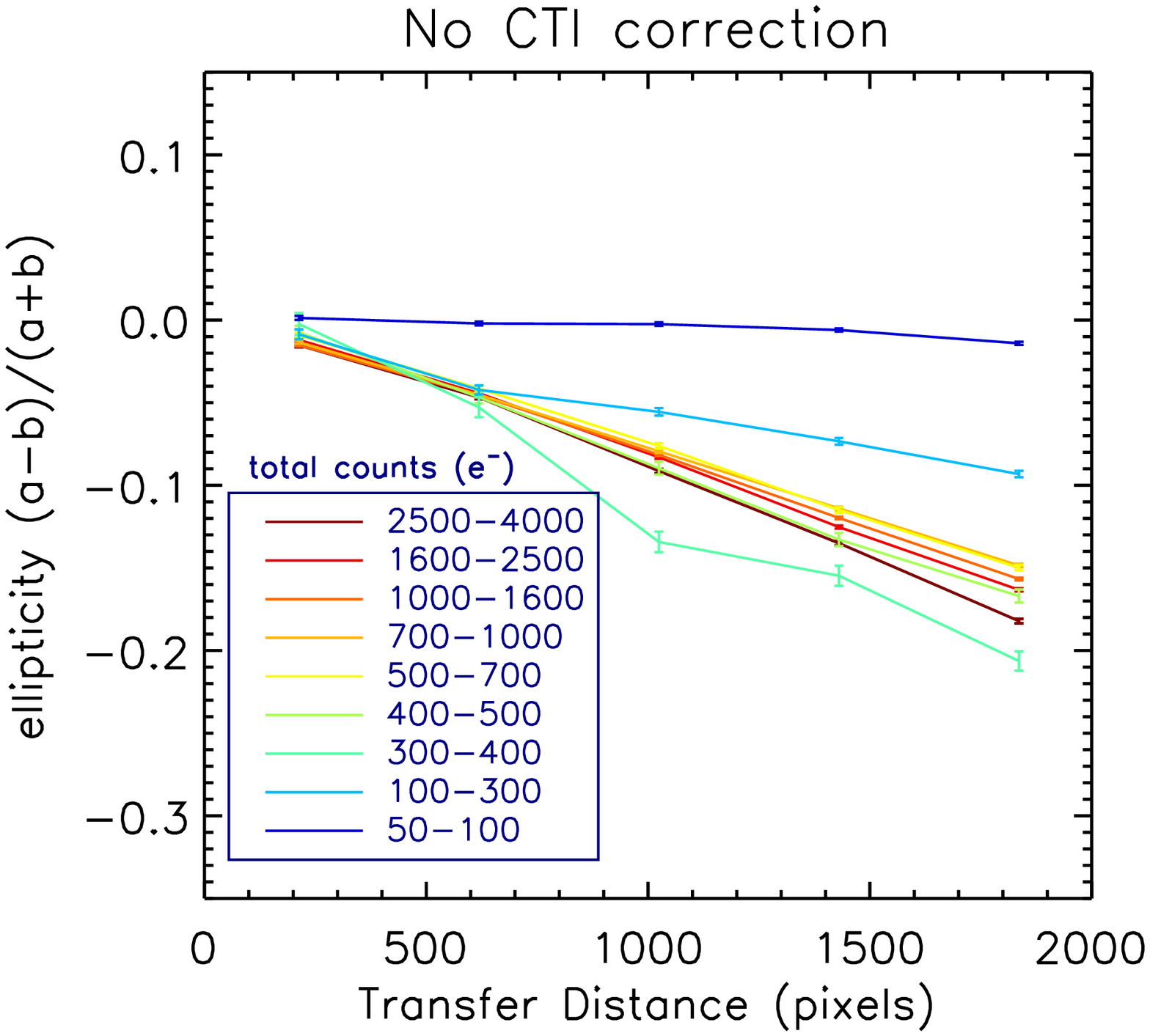}
\includegraphics[width=6cm]{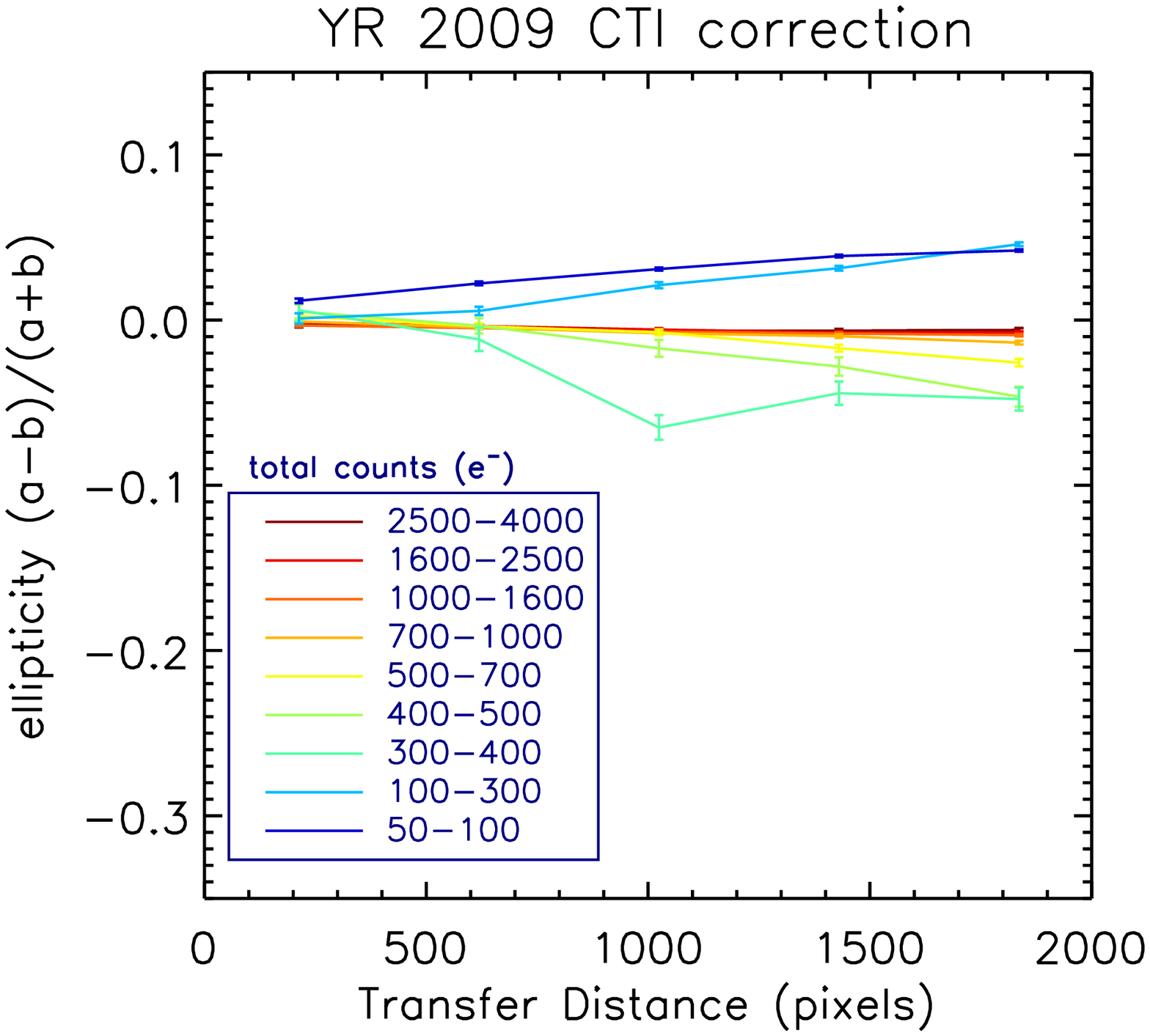}
\includegraphics[width=6cm]{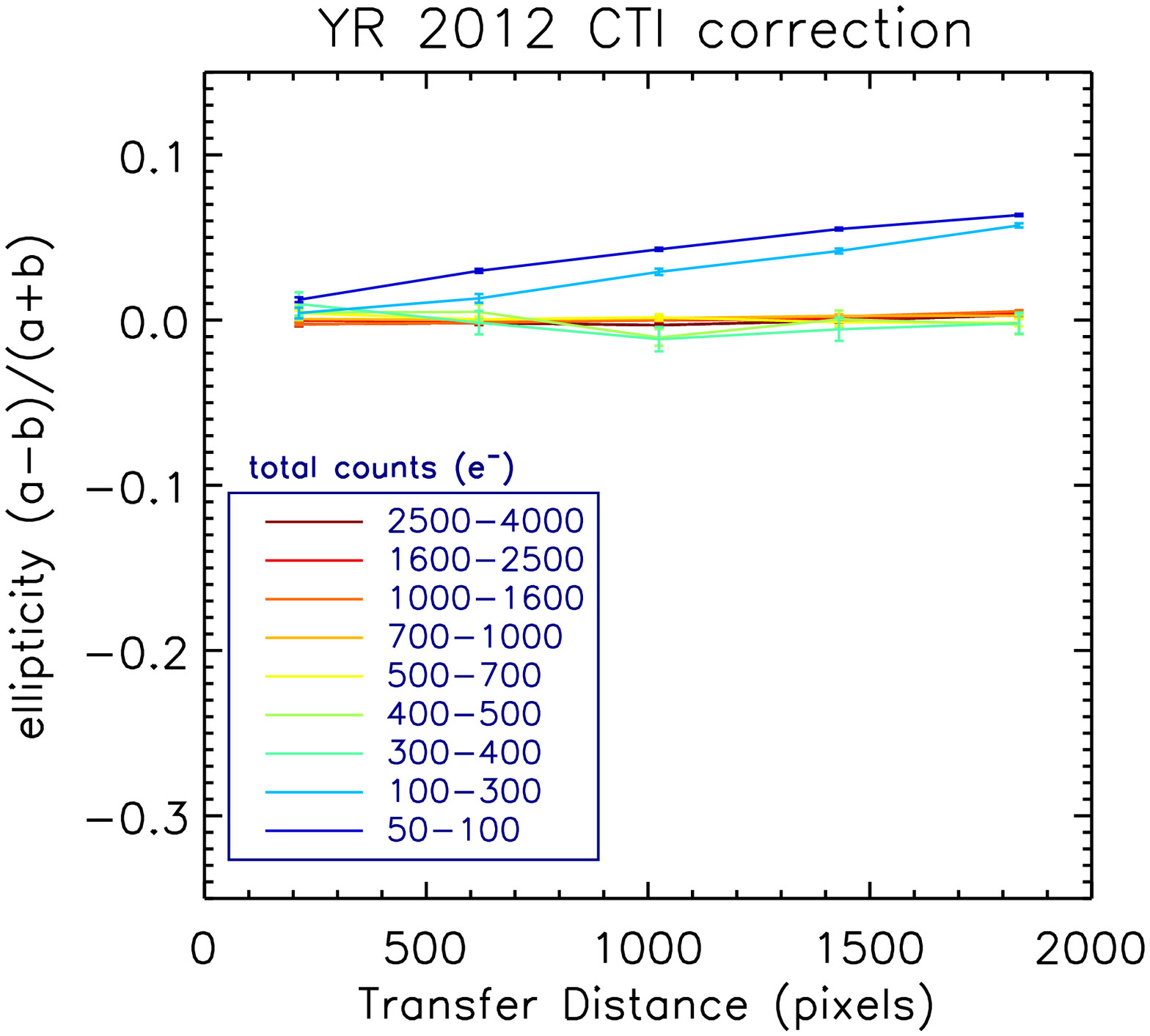}
\caption{CTI effects in the A520 ACS data measured from the
  ellipticity of sub-PSF features (SPFs) such as cosmic rays and
  warm pixels. The counts are calculated with the background 
  subtracted. The left plot shows the result from the uncorrected ACS
  images. Without any correction, severe charge trailings are present,
  and thus weak lensing based on the raw images will be non-negligibly
  biased. The middle panel displays the result when the old (Anderson
  \& Bedin 2010) pixel-based method is applied, which reduces the CTI
  artifacts.  However, the residual ellipticities indicate that the
  results are not yet satisfactory. The right panel shows the result
  when the latest (Ubeda and Anderson 2012) correction is used.
  The latest correction method successfully takes care
  of these residual CTI effects for the SPFs with counts greater than
  $\mytilde300$.  Both corrections tend to overcorrect the
  CTI effects for low-count objects ($\mytilde300$). This might be
  because the current pixel-based method does not take into account
  the low-flux-limit CTI-mitigation reported in Jee et al. (2009).
\label{fig_cti_ellipticity}}
\end{figure*}

\section{ANALYSIS}\label{section_analysis}

The key ingredient in any weak lensing analysis is the accurate
measurement of the shapes of the source galaxies. A number of
observational effects prevent us from simply using the observed
shapes. Instead we need to characterize these instrumental distortions
and correct for them. In addition to the usual correction for the PSF
(both size and anisotropy), the analysis of ACS data needs to account
for the trailing of charge due to CTI.
Since charge trapping happens every
time charges are transferred from one pixel to another, CTI is greater
for pixels farther from the readout register. After the trapped
charges in one transfer are released, a fraction of them remain
trapped during the next transfer. This cascading effect leads to
trails, leading to a change in the shapes of galaxies. Both photometry
and shape measurements are affected by CTI, although both are
sensitive to different species of traps.

Although this undesirable artifact happens in every CCD, it is an
especially serious concern for space telescopes where the CCDs are
subject to constant space radiation and the sky background is very
low. The number of defects increases roughly linearly with time, and
the A520 data taken in the Cycle 18 (2011 February and April) are
severely affected by this CTI problem. 

J12 discussed the potential impact of uncorrected CTI in the WFPC2
analysis. Although WFPC2 at the time of the observation of A520 was in
orbit longer than ACS at the time of the current observation, the
much smaller size of the readout distance (800 pixels vs. 2048 pixels)
makes the overall impact less severe. In addition, the direction of
the CTI trails differs for the three CCDs of WFPC2, which results in a
more effective mixing of any residual CTI pattern.

\subsection{Correction for CTI} \label{section_cti_impacts}

We measure the CTI utilizing warm pixels and compact cosmic rays
present in the same science data (Jee et al. 2009; Jee et al. 2011;
J12). These sub-PSF features (hereafter SPFs) suffer from the charge transfer
efficiency degradation but are not affected by the anisotropic PSF of
the instrument. Hence we can single out CTI and perform a statistical
analysis of their ellipticity as a function of charge transfer
distance and flux.

We quantify the elongation of SPFs due to CTI using the ellipticity defined as
\begin{equation}
e=\frac{a-b}{a+b}
\end{equation}
\begin{equation}
e_+= e \cos (2\theta) 
\end{equation}
\begin{equation}
e_\times= e \sin (2\theta) 
\end{equation}
\noindent
where $a$ and $b$ are the semi-major and -minor axis, respectively. 
$\theta$ is  the orientation of the ellipse. For the current CTI measurement, 
we choose the serial readout direction as our $x$-axis and the parallel
readout direction as our $y$-axsis.

The plot on the left panel of Figure~\ref{fig_cti_ellipticity} shows
the $e_{+}$ component of the SPF ellipticity as a function of a charge
transfer distance; because the $x$-axis is defined to be orthogonal to
the readout direction, the sign of $e_{+}$ becomes negative for CTI
trails.  Different colors represent different ranges of flux counts
(after background being subtracted).  Several features are worth
noting in comparison with previous work. First, the CTI is still
linear with transfer distance, as was observed in our previous studies
(Jee et al. 2009; Jee et al. 2011; Hoekstra et al. 2011). Second, even
the brightest SPFs are severely affected by CTI. Compared to our 2009
data, the slope of the SPFs at the flux range 2500-4000 $e^{-}$ has
increased by more than a factor of two, showing that the farthest
($\mytilde2000$ pixels) SPFs suffer from a net ellipticity
distortion of 0.2 whereas it was at the $\delta e\sim0.1$ level for
the data taken in the year 2009.  Third, the flux-dependence of the
slope is more complicated. In Jee et al. (2009; 2011), we observed
that the CTI slope becomes more negative for decreasing flux until it
sharply turns around at $\mytilde300~e^{-}$ (see Figure 31 of Jee et
al. 2011), which would affect objects fainter than $F814W \sim 27$.
However, the A520 images show that the slope becomes less negative for
decreasing flux initially and then suddenly turns around at
$\mytilde400~e^{-}$. At the faintest limit, it turns around again, and
the CTI is mitigatied once more. 

Anderson \& Bedin (2010) developed a so-called pixel-based CTI
correction method, and their standalone script {\tt PixCteCorr} is
publicly available and can be applied to regular {\tt FLT} files. The
results shown in the middle panel of Figure~\ref{fig_cti_ellipticity}
are obtained from these images (hereafter we refer to this method as
the Y2009 model) when we repeat the above experiment.  The Y2009 model
reduces the CTI effects substantially, and the performance is
excellent especially in the brightest regime ($700-4000
~e^{-}$). Nevertheless, the model does under-correct the CTI in the
intermediate flux range ($300-700~e^{-}$). However, the most
interesting feature is the behavior of the SPFs at the faintest end
($50-300~e^{-}$).  The CTI slopes are $positive$ in this flux range
because the model $overcorrects$ the CTI. This also serves as proof of
the CTI mitigation at the faint limit first reported in Jee et
al. (2009) and later supported by Schrabback et al. (2010).

\begin{figure*}
\begin{center}
\hbox{%
\includegraphics[width=0.32\hsize]{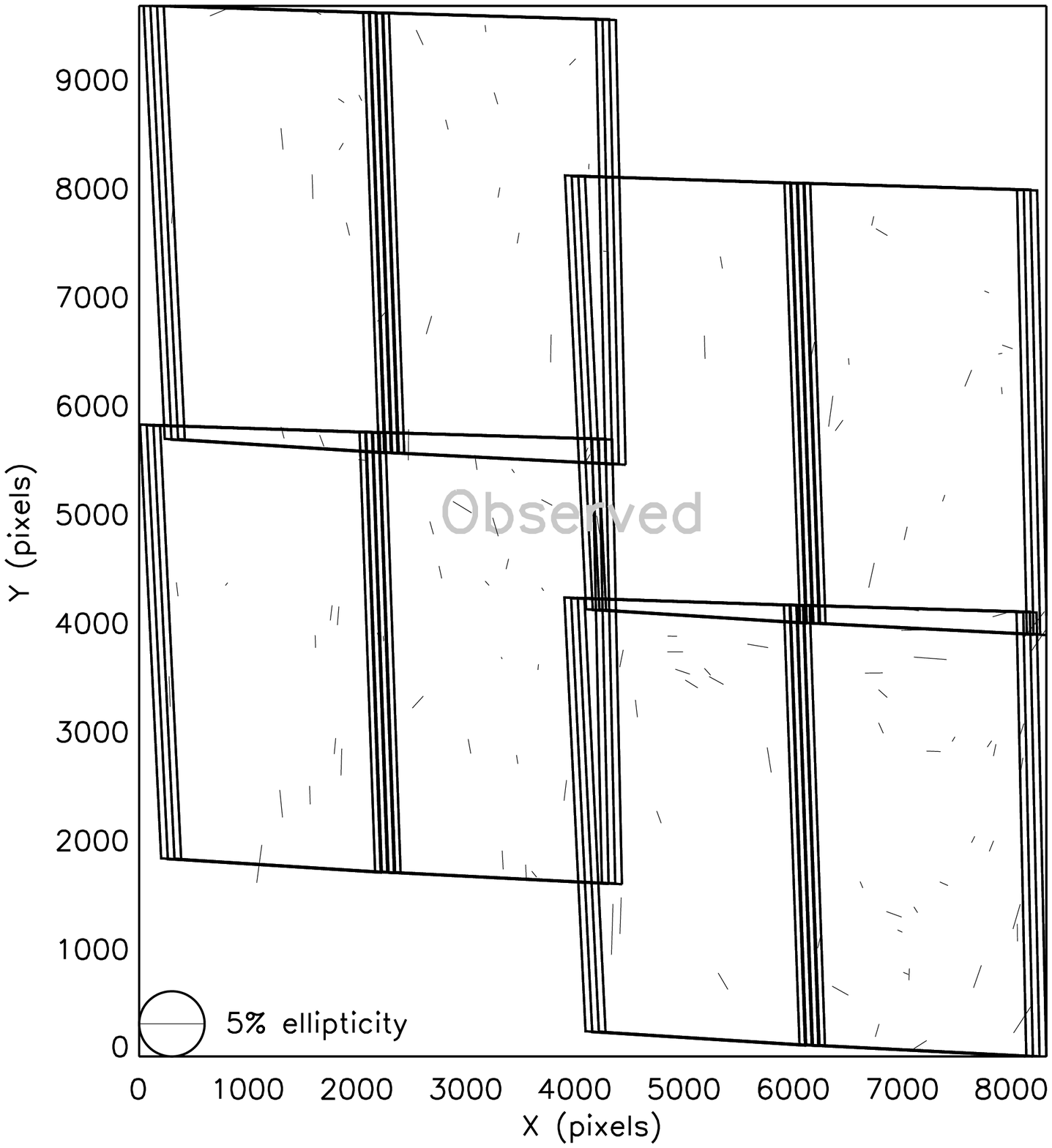}
\includegraphics[width=0.32\hsize]{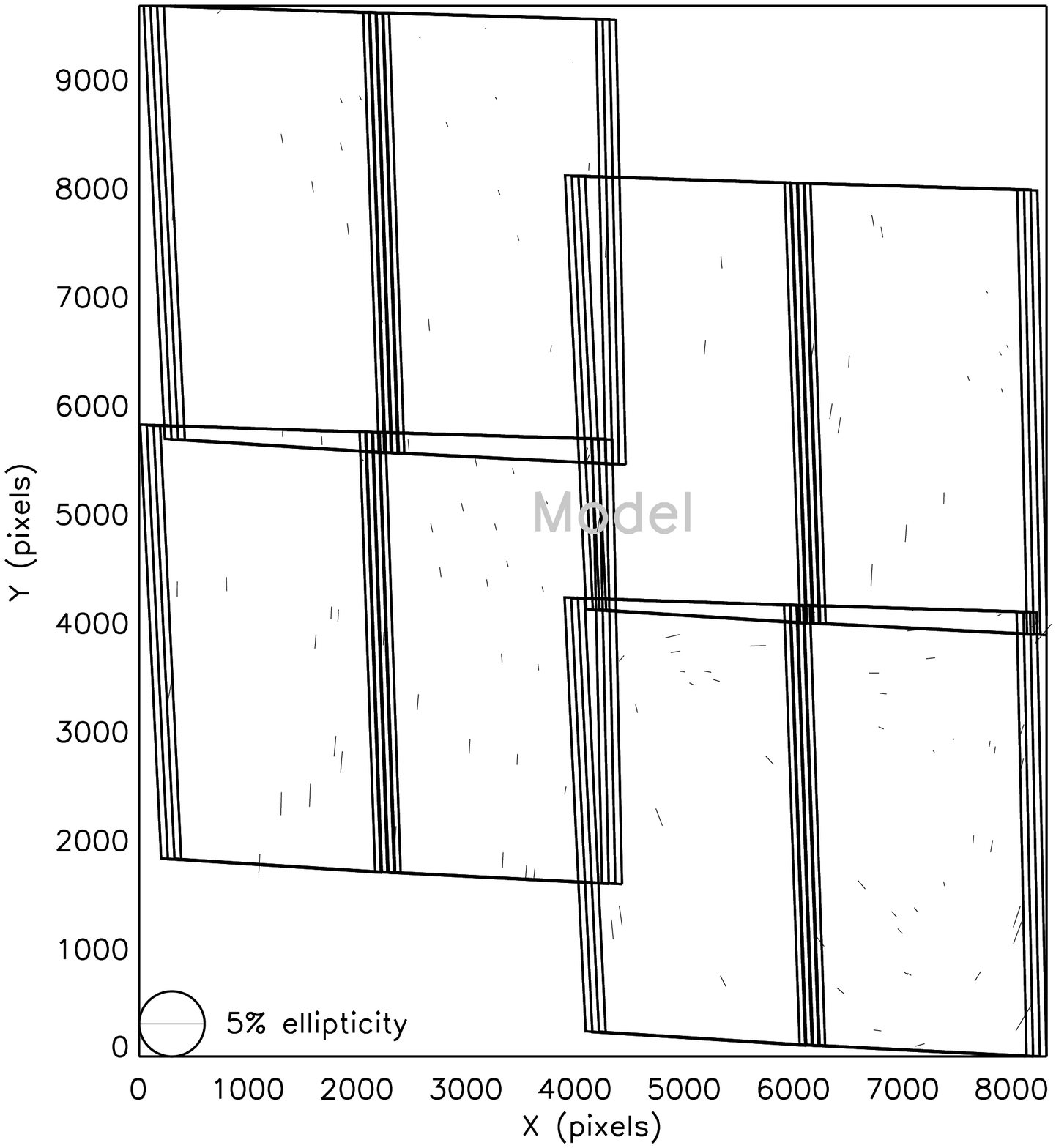}
\includegraphics[width=0.32\hsize]{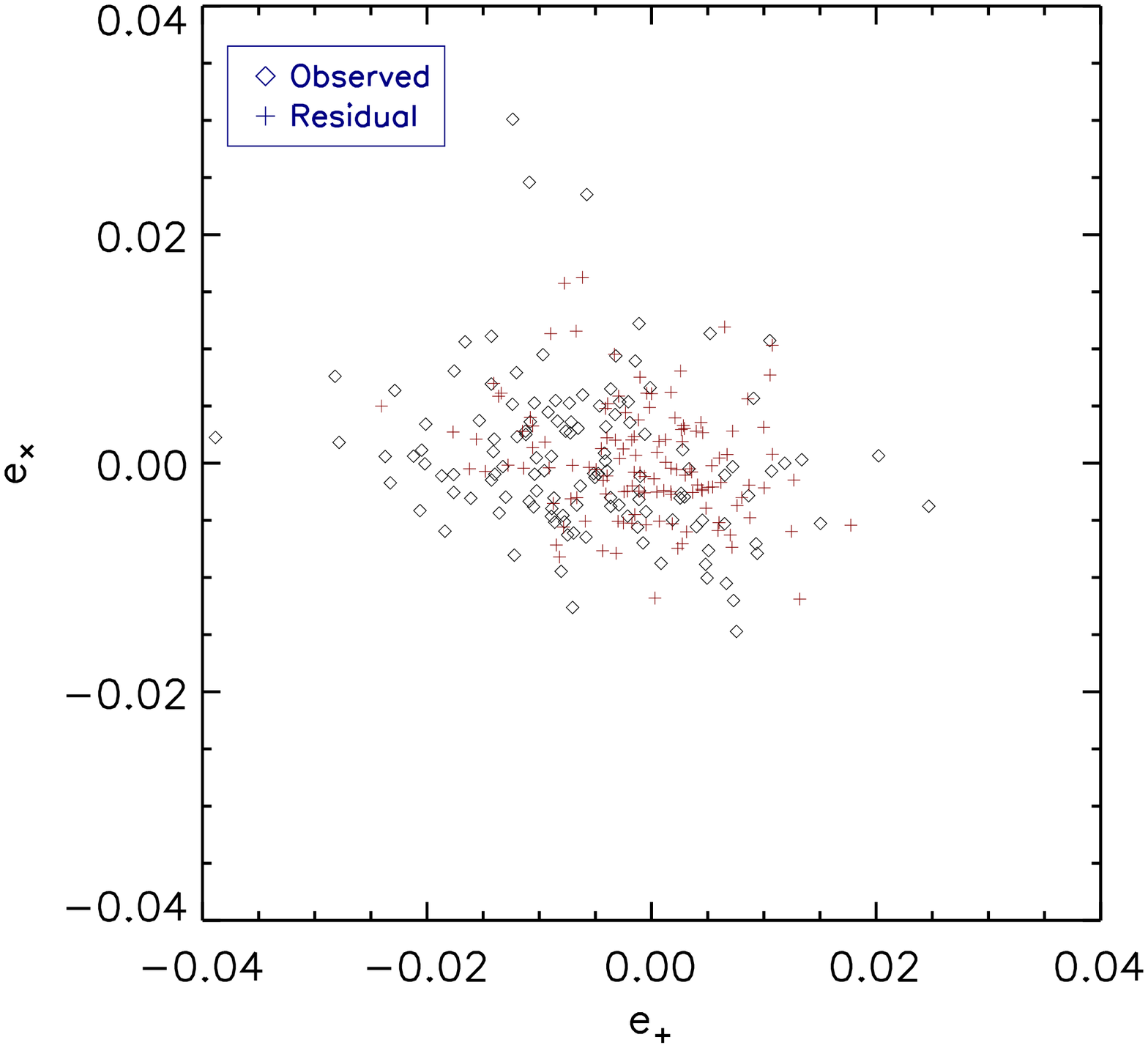}}
\end{center}
\caption{PSF reconstruction in the F814W images. The left panel shows
  the observed ellipticity pattern of the stars whereas in the middle
  panel we display the ellipticity of the model PSF derived from
  globular cluster fields. The sticks illustrate the direction and
  magnitude of the ellipticity of the PSFs.  The solid lines represent
  the observation footprints. 
  The plot in the right panel displays the residual ellipticity of the stars.
\label{fig_psf_model}}
\end{figure*}

A new CTI correction method has been proposed by Ubeda and Anderson
(2012) and it is now part of the default STScI pipeline. This method
(hereafter the Y2012 model) is an important improvement over the
earlier method in that it includes both time and temperature
dependence. In addition, Ubeda and Anderson (2012) state that the
performance in the low flux regime has been significantly improved.
We display the results obtained from this new correction in the right
panel of Figure~\ref{fig_cti_ellipticity}.  It is clear that the
undercorrection problems (e.g., see the slope for the range
$300-400~e^{-}$ in the middle panel) seen in the old model nicely
disappear in this case. However, unfortunately, the overcorrection
problem at the faint limit still remains (perhaps becoming even slightly
worse). Nonetheless the performance of this correction is the best and we
use the Y2012 CTI-correction algorithm for our actual weak lensing
analysis.

The remaining concern is the treatment of the overcorrection problems
at the faint limit. The flux range $50-300~e^{-}$ corresponds roughly
to $F814W=27-28.5$. However, this blind conversion is not accurate
mainly because galaxy profiles are much less steep than those of the
SPFs.  Thus, any naive attempt to merely correct the ellipticity of
these sources at the catalog level will likely fail.  As in Jee et
al. (2011), we correct for these residual CTI-correction errors by
modifying our model PSF by treating the distortion of the PSF due to
CTI as an additional convolution. This assumption should hold better
in the current study than in Jee et al. (2011) because most
significant CTI-effects are already fixed by the pixel-based method
(thus, the residual CTI effect on PSF can be approximated by slightly
stretching the PSF). We find that about 15\% of our source galaxies need this
additional correction. Because the residual correction is small and
the ellipticities of these faint galaxies are already down-weighted,
the impact due to the imperfection of our residual correction on weak
lensing analysis is negligibly small 

Another method to address the overcorrection would be simply to discard
the sources at the faint end that are likely to be over-corrected by
the Y2012 algorithm. We verify that our weak-lensing result obtained with this
scheme is highly consistent with those from the full catalog, albeit with a slightly
higher noise level.

\subsection{PSF Model}

Although the PSF of ACS is small, a careful effort must be made to
model and remove the smearing effect of the PSF.  In particular, when one
desires to utilize sources near the 5$\sigma$ detection limit and the size of the PSF, the
complex spatial variation of the ACS PSF should be fully considered to
avoid bias due to the PSF anisotropy (the sizes of these faint sources
are comparable to that of the PSF).  

About 10-20 high S/N ($>20$) stars  are found in the individual
exposures of the A520 data. Since the ACS PSF is both time- and
position-dependent, it is impossible to use this small number of stars
to derive a reliable PSF model at the location of galaxies across the
field.  Therefore, we utilize the PSF library of Jee et al. (2007a)
constructed from dense stellar fields. The PSF pattern of ACS is
repeatable (Jee et al. 2007a), and thus can be estimated for each
exposure by measuring the shape properties of the stars in the
A520 data and comparing them with those from the PSF
library. We keep track of the model PSFs at the locations of source galaxies
in individual exposures
to compute the final PSF on the stack image. This requires a rigorous
propagation of the image stacking history including offset, rotation,
and weight applied to individual exposures. We refer readers to our
previous publication (e.g., Jee et al. 2011) for details.

Figure~\ref{fig_psf_model} compares the PSF ellipticity pattern of the
stars and the model PSFs in the A520 F814W images, where we measure
shears. The comparison shows that our PSF model reasonably mimics the
observed PSF pattern. Any significant registration error is supposed to create 
spurious stellar ellipticities not observed in the model. We do not find
any hints of such a large discrepancy.
The residual PSF ellipticity
rms per component is small ($\mytilde0.006$), and thus the impact of
the residual anisotropy on shear is negligible.

\subsection{Shape Measurement and Cluster/Source Galaxy Separation} 
\label{section_source_selection}

Our ellipticity is defined by $(a-b)/(a+b)$, where $a$ and $b$ are the
semi-major and minor axes, respectively\footnote{An alternative
  definition of shape using $(a^2-b^2)/(a^2+b^2)$ is often used in the
  literature. One should remember that this so-called polarisation
  needs to be divided by 2 to obtain an estimate of the shear.}.  We
determine the ellipticity of an object by fitting a PSF-convolved
elliptical Gaussian.  

Because the elliptical Gaussian is not the best representation of real
galaxies, it is important to correct for this ``underfitting"
(Bernstein 2010) together with other shear calibration issues such as
the dilution of the signal by noise and spurious sources. Our internal
shear calibration utilizing the Hubble Ultra Deep Field (Beckwith et
al. 2006; HUDF) data shows that the average correction factor is
$\sim11$\%. This average value is greatly influenced by low-surface
brightness galaxies at $F814W>26$, which require a slightly larger ($\sim14$\% on average) correction
factor, but still contain a useful lensing signal. This S/N-dependent
correction is often called noise bias (Melchior \& Voila 2012; Refregier et al. 2012), and
is an important factor affecting cosmic shear results.
In this paper, we multiply a S/N-independent average
correction factor 1.11 to our galaxy ellipticity. However,
we verify that analysis using a S/N-dependent correction scheme
yields virtually indistinguishable weak-lensing results.
Note that this
multiplicative bias affects the amplitude of the lensing signal, but should
not affect features in the mass map. 
To maximize the S/N of the lensing signal, the ellipticity of galaxies
must be properly weighted by taking into account both the source
ellipticity distribution and measurement errors. We
employ the following simple inverse-variance weighting scheme:

\begin{equation}
\mu_i  = \frac {1} { \sigma_{SN}^2 + (\delta e_i)^2} \label{eqn_shear_weight},
\end{equation}

\noindent where $\sigma_{SN}$ is the dispersion of the source
ellipticity distribution ($\mytilde0.25$ per component for the A520 data), and
$\delta e_i$ is the $i^{th}$ galaxy's ellipticity measurement error
per component. 

For bright galaxies (F814W$<24$), we use both F435W-F606W and
F606W-F814W colors to select the A520 members utilizing the criteria
of C12 (i.e., objects inside the quadrilateral in Figure~1 of C12).
This gives a total of 447 objects, and we use them to estimate the
cluster luminosity. For the source galaxy sample, we select galaxies
between $22<F814W<27.5$ and remove the bright (F814W$<24$) cluster
members. We do not attempt to remove cluster members at F814W$>24$
because the above color-based selection is not efficient in this
regime (in \textsection\ref{section_redshift} we demonstrate that in
fact the contamination from faint cluster members is negligible at
F814W$>24$).  Stars are identified utilizing both half-light radius
and shape measurement results.  Because delta-function-like features
should arise for point sources after deconvolution, we require the
minimum semi-minor axis to be 0.4 pixels to prevent accidental
inclusion of stars. The maximum allowed ellipticity error per
component is set to 0.25.  The total number of sources after these
cuts is 4,932, giving us a source density of $\mytilde109$ galaxies
per sq. arcmin.  C12 quotes $\mytilde56$ galaxies per sq. arcmin in
their ACS weak lensing analysis of A520. 

According to
Equation~\ref{eqn_shear_weight}, the uncertainty of the shear is given
by
\begin{equation}
\sigma_{\gamma} = \sqrt { \frac{1}{\Sigma \mu_i} }.
\end{equation}
\noindent On the other hand, if no weighting scheme is used, the
shear uncertainty is simply:

\begin{equation}
\sigma_{\gamma} = \frac{\sigma_{SN}} {\sqrt{n}} .
\end{equation}
\noindent We define the effective number by equating the last two equations and obtain

\begin{equation}
n_{\rm eff} = \sum \frac{\sigma_{SN}^2} {\sigma_{SN}^2 + (\delta e_i)^2}.
\end{equation}
\noindent Therefore, the effective number is always smaller than the
actual number of sources.  We estimate the effective source density
to be $\mytilde96$ per sq. arcmin. The corresponding rms shear is
$\mytilde0.026$ per sq. arcmin, which is $\mytilde$28\% smaller than
the value quoted by C12 ($\mytilde0.036$ per sq. arcmin).

\begin{figure}
\begin{center}   
\includegraphics[width=\hsize]{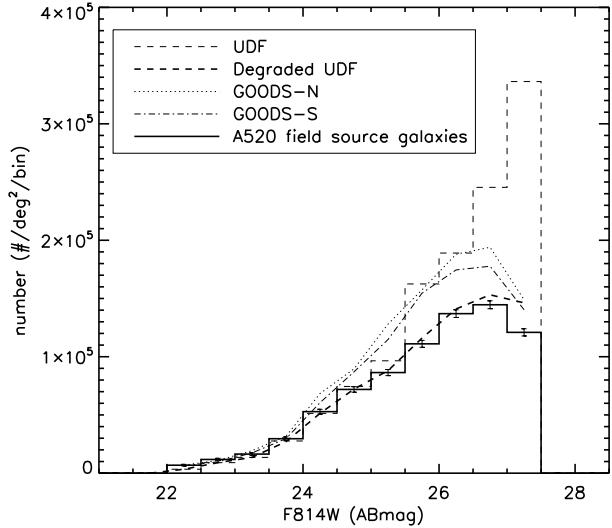}
\end{center}
\caption{Magnitude distribution of source galaxies in the A520
  field. We apply the identical selection (after due color
  transformation) criteria to the UDF and GOODS data and show the
  resulting magnitude distributions for comparison. For relatively
  bright source galaxies ($F814W \lesssim24.5$), the magnitude
  distributions from the four fields agree nicely, indicating that the
  contamination from unidentified cluster members (i.e., missed by our
  color selection) is negligible in this regime. Because both UDF and
  GOODS images are much deeper than the A520 one, it is difficult to
  directly compare the distribution at $F814W
  \gtrsim24.5$. Nevertheless, when we degrade these control fields in
  such a way that the noise levels become comparable, we observe a
  good agreement in this faint regime, too, up to the sample
  variance. The displayed error bars include only Poissonian noise.
\label{fig_mag_dist}}
\end{figure}

\subsection{Redshift Estimation of Source Population} \label{section_redshift}

Since we do not apply any color cut for galaxies fainter than
F814W$\sim24$, it is important to estimate the level of potential
contamination in our source catalog carefully and to propagate it to
our redshift estimation.  To address this issue, we utilize the Coe et
al. (2006) HUDF photo-z catalog, the 2004 STScI release of the HUDF
images (Beckwith et al. 2006), and the 2008 STScI release of the Great
Observatories Origins Deep Survey (GOODS; Giavalisco et al. 2004)
images. We apply the same source selection criteria to our galaxy
catalog of the HUDF and GOODS data and compare their magnitude
distributions with those from the source population in Figure~\ref{fig_mag_dist}.

The comparison shows no excess of galaxies in the A520 source catalog
with respect to the magnitude distribution computed from the above
reference fields.  At the faint end, the number densities of the
galaxies in the reference fields are somewhat higher than those in the
A520 field simply because of their increased depth. To enable a fair
comparison, we degrade the reference images to match the noise level
of the A520 images. The resulting distribution from the UDF is in
good agreement with that of the A520 source galaxies. 
Our test with the GOODS
images also confirms that the cluster galaxy contamination is negligible.

In order to scale our lensing signal properly, we must estimate the $
\beta$ parameter defined as:

\begin{equation}
\beta  = \left < \mbox{max} \left (0,\frac{D_{ls}}{D_s} \right ) \right >, 
\end{equation}

\noindent where $D_{ls}$ and $D_s$ are the angular diameter distances
between the lens and the source, and between the observer and the
source, respectively.  This $\beta$ parameter determines the critical
surface mass density $\Sigma_c$ of the cluster given by

\begin{equation}
\Sigma_{\rm crit} = \frac{c^2}{4 \pi G D_l \beta }, \label{eqn_sigma_c}
\end{equation}

\noindent where $c$ is the speed of light, $G$ is the gravitational
constant, and $D_l$ is the angular diameter distance to the lens.
Compared to high-redshift clusters, the weak lensing mass of A520 at
$z=0.2$ is not as sensitive to the source redshift. However, we take
care to apply the same source selection criteria to the UDF data.

Using only the color and magnitude criteria gives $ \beta =0.75$,
which corresponds to an effective source plane at $z_{\rm eff}\sim1$.
This value is slightly biased high because there are fewer galaxies at
the faint limit in the A520 data than in the UDF image.  In addition,
the ellipticities of these faint galaxies are down-weighted when we
estimate shear.  Considering both effects, the revised estimate
becomes $\beta =0.73$ or $z_{\rm eff}\sim0.85$.  The resulting critical
mass density $\Sigma_{\rm crit}$ is $3.35\times10^3 M_{\sun} \mbox{pc}^{-2}$.

C12 quote $\Sigma_{\rm crit}=3.6\times10^3 M_{\sun} \mbox{pc}^{-2}$ for their
weak lensing analysis. This higher surface mass density implies that their source
redshift is lower than ours, which is consistent with the fact that
our source catalog contains more faint galaxies than theirs\footnote{We
note that C12 estimate the critical surface mass density of J12 to be
$\Sigma_{\rm crit}\simeq4.1\times10^3 M_{\sun} \mbox{pc}^{-2}$ using
$\beta=0.64$ quoted in J12.  However, substituting this $\beta$ value
into equation~\ref{eqn_sigma_c} yields $\Sigma_{\rm crit}\simeq3.8\times10^3
M_{\sun} \mbox{pc}^{-2}$; the difference in the assumed cosmological
parameters between J12 and C12 makes only a negligible change in the
conversion of $\beta$ to $\Sigma_{\rm crit}$.}.

\begin{figure}
\begin{center}
\includegraphics[width=\hsize]{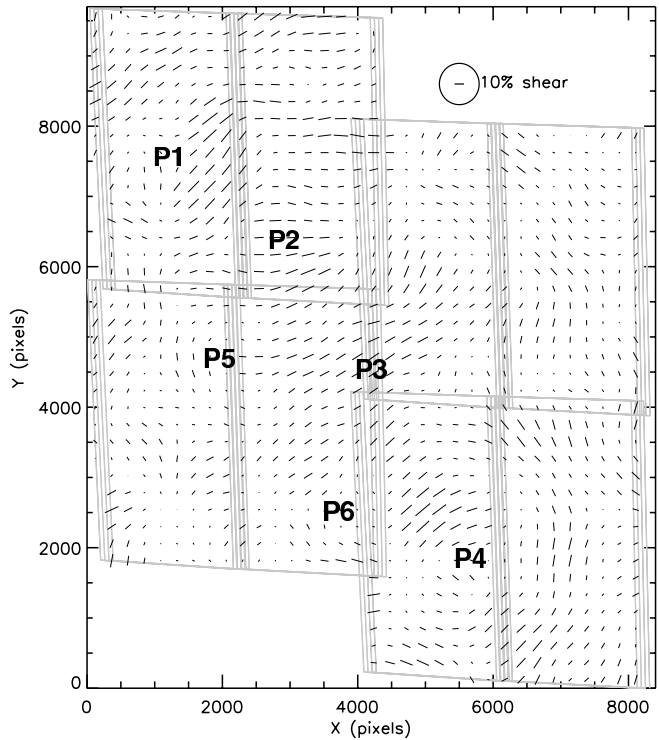}
\end{center}
\caption{Smoothed ellipticity distribution of source galaxies. 
The
  ``whisker" plot is produced by convolving the ellipticities with a
  Gaussian kernel. The diameter of the circle represents the FWHM
  ($30\arcsec$) of the convolution kernel while the stick inside this
  circle shows a 10\% horizontal shear.  Clear correlation of source
  galaxy ellipticity is seen. The approximate locations
  of the substructures reported in J12 are annotated with P1-P6.
\label{fig_whisker}}
\end{figure}

\section{Mass reconstruction} \label{section_mass_reconstruction}

An important application of weak gravitational lensing is that the
observed shear signal can be used to reconstruct the projected mass
density. The smoothed shear field presented in
Figure~\ref{fig_whisker} shows a coherent pattern around the main
galaxy overdensities. This can be related directly to the convergence
map $\kappa(\bvec{x})=\Sigma(\bvec{x})/\Sigma_{\rm crit}$ through

\begin{equation}
\kappa (\bvec{x}) = \frac{1}{\pi} \int D^*(\bvec{x}-\bvec{x}^\prime)
\gamma (\bvec{x}^\prime) d^2 \bvec{x} \label{k_of_gamma}.
\end{equation}

\noindent where $D^*(\bvec{x} )$ is the complex conjugate of the
convolution kernel $D(\bvec{x} ) = - 1/ (x_1 - \bvec{i} x_2 )^2$ and
$\gamma(\bvec{x})$ is the complex representation of gravitational shear.  

\begin{figure*}
\begin{center}
\includegraphics[width=0.42\hsize]{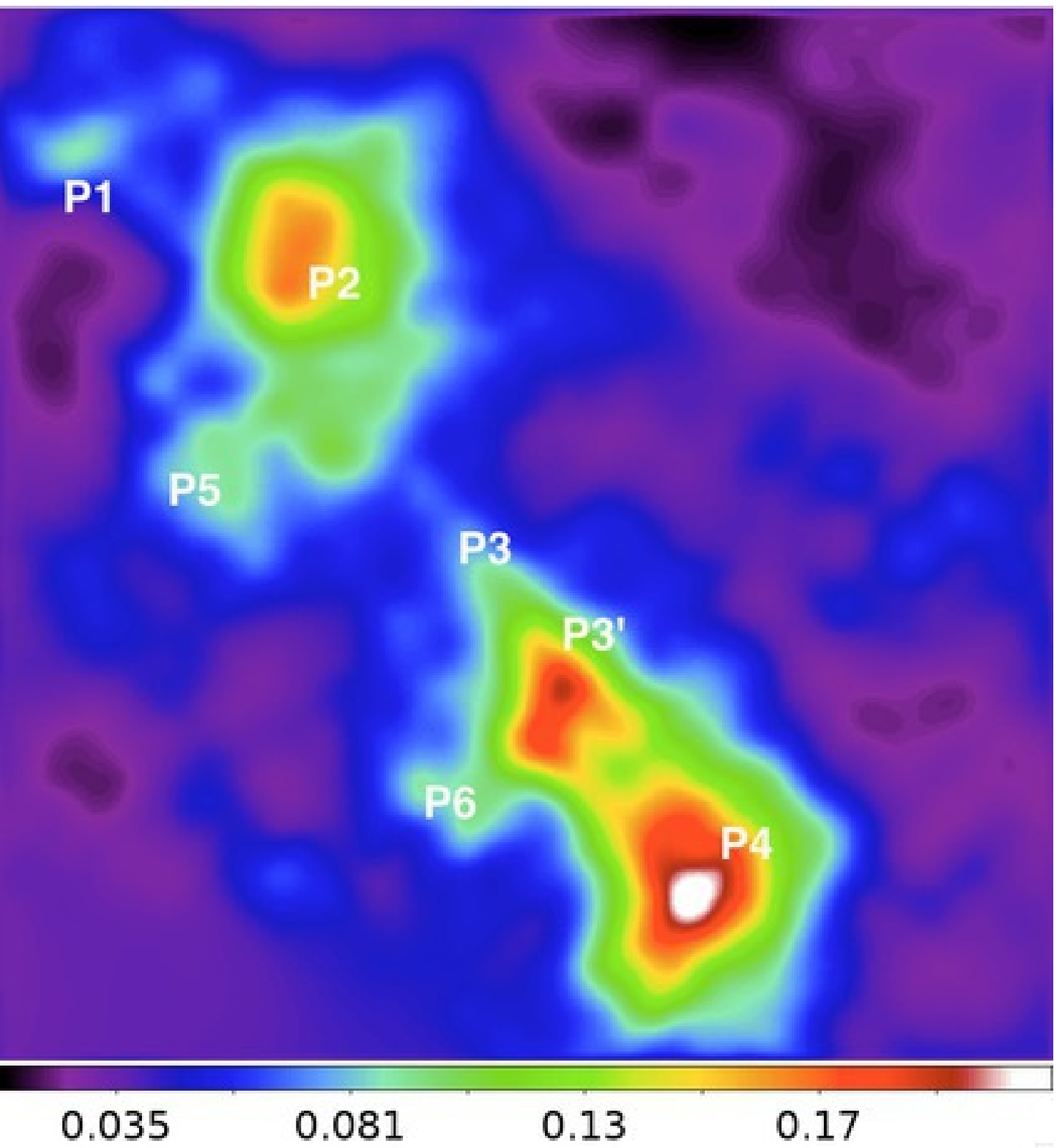}
\includegraphics[trim = 0mm 9mm 0mm 0mm, width=0.53\hsize]{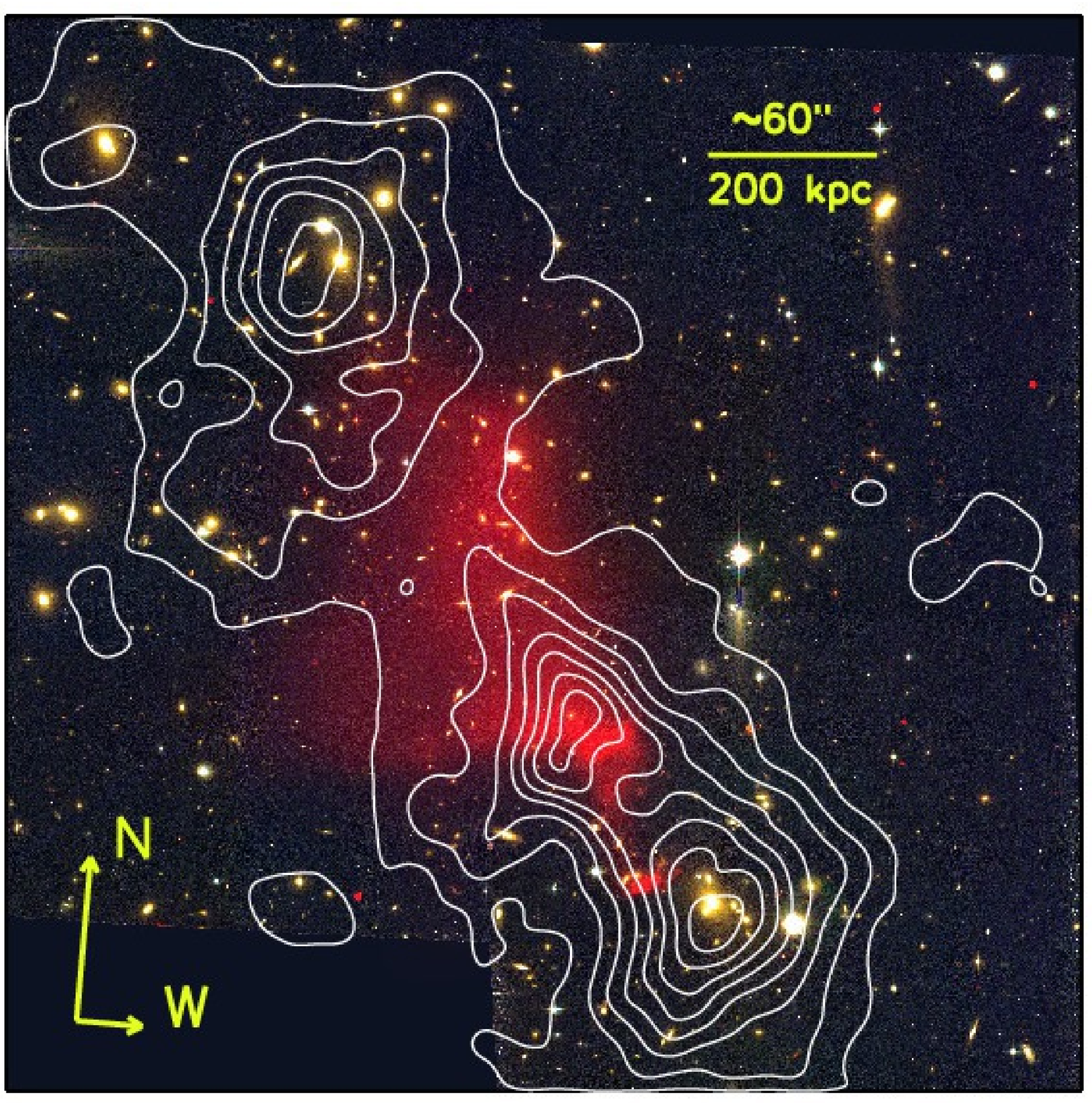}
\end{center}
\caption{Maximum-entropy-regularized mass reconstruction of A520. The
  left panel shows the density distribution using the color scheme
  shown at the bottom.  We annotate the location of the substructures
  reported in J12 as P1-P6.  On the right panel, we overlay the mass
  contours on the ACS color composite image. The intensity of the
  X-ray emission observed by Chandra is represented in red.
\label{fig_massmap_contour}}
\end{figure*}

A better result is produced if we do not pre-smooth the ellipticities,
but instead let the strength of the shear signal determine the local
smoothing scale. One such method is the maximum-entropy-regularized
mass reconstruction first introduced by Seitz et al. (1998). In this
study, we use the Jee et al. (2007b) implementation of the
method. Figure~\ref{fig_massmap_contour} shows the result. A similar
result is obtained when we use equation~\ref{k_of_gamma}, although the
map becomes noisier near the edges.

\begin{deluxetable*}{lcccc}
\tabletypesize{\scriptsize}
\tablecaption{Mass Properties of Substructure ($r<150$~kpc)}
\tablenum{1}
\tablehead{\colhead{Substructure} & \colhead{$\alpha, \delta$} & \colhead{$\Delta \alpha, \Delta \delta$} & \colhead{Projected Mass} & \colhead{Gas Mass}   \\
                    \colhead{}   & \colhead{ ($^h~^m~^s$, $\degr~ \arcmin~ \arcsec$)} & \colhead{ ($\arcsec,\arcsec$)}& \colhead{($h_{70}^{-1} 10^{13} M_{\sun}$)}  & \colhead{$(h_{70}^{-5/2} 10^{13} M_{\sun})$}  \\}
\tablewidth{0pt}
\startdata
P1  &  (04 54 20.76,  +02 57 38.4)   & (5.3,4.1) &                			$2.10\pm0.43$ &    $  <0.23$\\
P2  & (04 54 15.02,  +02 57 09.2)    & (4.0,4.9) &                			$4.05\pm0.28$ &    $  <0.40 $\\
P3 (old centroid) & (04 54 11.07, +02 55 35.3) &  - &               $3.35\pm0.34$ &                 $  <0.74$ \\
P3' (new centroid) & (04 54 07.51, +02 54 41.3) & (5.1,6.2) &             $3.94\pm0.30$ &   $  <0.85$ \\
P4  & (04 54 04.32,  +02 53 51.0)  & (3.0,4.4)   & 	       			$4.23\pm0.28$ &   $ <0.34$ \\
P5  & (04 54 16.53,  +02 55 26.7)  & (5.4,8.2)   &	      			$2.93\pm0.39$ &   $  <0.21$ 
\enddata
\tablecomments{The positional uncertainty is estimated from bootstrapping.
The mass uncertainties are evaluated from 1000 Monte-Carlo realizations.  
The gas mass is derived using Cauchy-Schwartz method in M07 based on the data set ObsID 9426 (110 ks)
for P3$^{\prime}$ and ObsId 528 (38 ks) for the other peaks. 
The Cauchy-Schwarz method is a model-independent means of deriving an {\it upper limit} on the gas mass 
column in a given region of the sky. The method requires an estimate of the maximum length of the 
cluster gas along the line of sight that contributes 99\% of the emission. We refer to M07 for details; 
here we use an updated maximum column length of 4 Mpc (instead of 2), which is a more conservative 
estimate given the total mass of Abell 520. As a result, our gas mass upper limits are more conservative, 
being higher than M07 by $\sqrt{2}$. The given values are 90\% confidence level upper limits.
}
\end{deluxetable*}

The substructures seen in this ACS analysis are in general similar to
those in the WFPC2 results of J12 and their locations are denoted as P1-P6; their
coordinates are listed in
Table~1.  However, one of the important differences is the location of
the dark peak. The current centroid (P3$^{\prime}$) is about
$1\arcmin$ shifted to the southwest compared to the one in J12 (P3). Although
mass reconstructions using different imaging data can sometimes result
in small offsets in the positions of the mass peaks, it is unusual to
observe a shift as large as $\mytilde1\arcmin$. 

Another noteworthy difference between our ACS and WFPC2 results is the
strength of the substructure P4. In our ACS result, P4 is the
strongest mass peak in the A520 field whereas it appears as a minor
(much weaker than P3) clump in our WFPC2 analysis. However, this
difference can arise from the incomplete coverage of the region by the WFPC2 observation.
In addition, we find that the ACS
images reveal many new arclets around P4, which all contribute to the
significance of the peak in mass reconstruction. We note that the
inferred projected mass of P4 is consistent with the value reported in
J12.

\begin{figure*}
\begin{center}
\includegraphics[width=0.8\hsize]{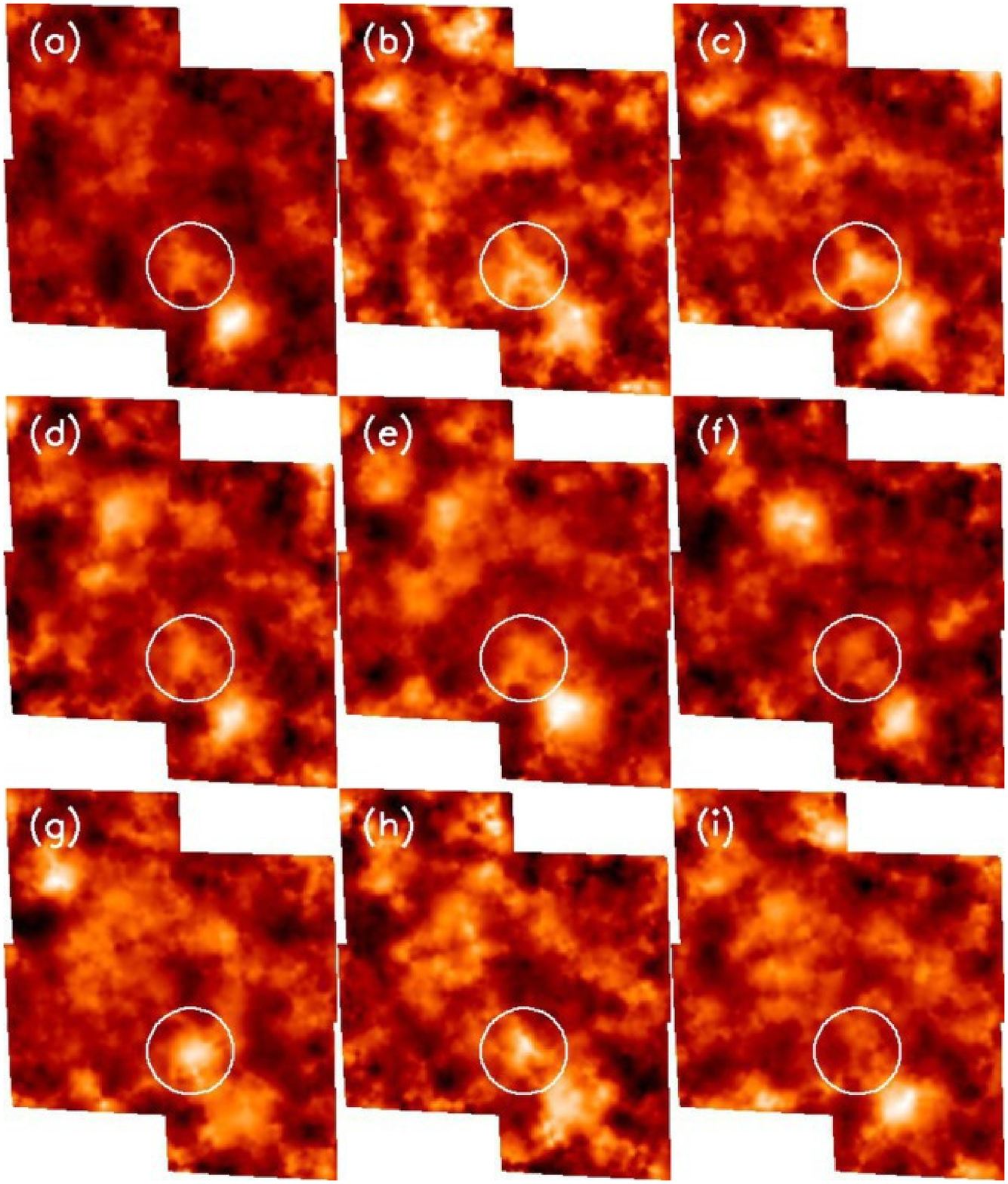}
\end{center}
\caption{Bootstrapping test of the A520 mass reconstruction. 
Mass reconstruction is performed with bootstrap-resampled 
source galaxies. The test provides a measure to
examine the statistical significance of the A520 substructures. 
We use the {\tt FIATMAP} code to generate 1000 results. 
Here we display nine random results. 
The circle denotes the approximate location of the dark peak.
\label{fig_bootstrap}}
\end{figure*}

\begin{figure}
\begin{center}
\includegraphics[width=8cm]{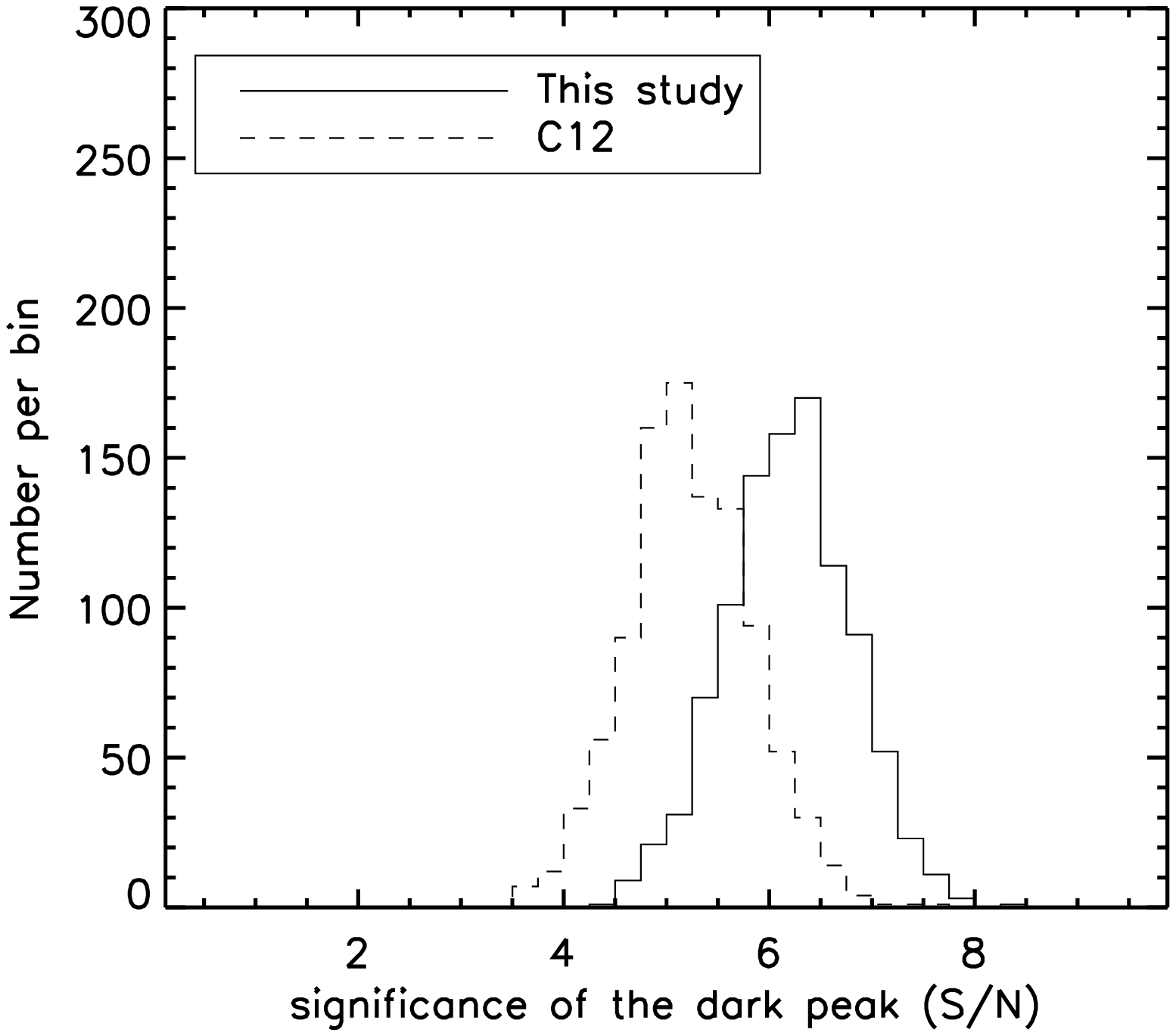}
\end{center}
\caption{Significance of the dark peak measured from our 1000 bootstrap runs.
The signficance is measured using a $r=150$kpc aperture against the background,
which is determined within each mass reconstruction field. See text for details.
\label{fig_sig_darkpeak}}
\end{figure}

\subsection{Centroid and Significance of the Dark Peak in A520}

We perform a bootstrapping analysis to measure both the significance and the positional
uncertainty of the substructures. Utilizing the fast Fischer \& Tyson
(1997) implementation {\tt FIATMAP} of the KS93 algorithm, we generate 1000
realizations. In Figure~\ref{fig_bootstrap},  we display nine random samples of the bootstrapped 
mass reconstructions. The circle denotes the approximate location of the dark peak

We measure the centroids using the first moments
weighted by a circular Gaussian, whose FWHM matches the size of the
substructure.  The results are displayed in Table~1.  The mean positional uncertainty
is small ($\mytilde5\arcsec$), and thus 
we conclude that the large shift of the dark peak centroid between the current
study and J12 is not caused by noise in the mass reconstruction. 
We discuss a number of possible explanations in
\textsection\ref{sec_shift}.  Nevertheless, we note that the 
aperture mass within $r=$150 kpc centered on P3 (old centroid) is still consistent
with the WFPC2 value (see \textsection\ref{section_mass_estimation}).
In other words, there is significant dark mass present in projection
about P3.

Quantifying the significance of the P3$^{\prime}$ substructure requires us to determine a
reasonable baseline in the absence of the dark peak.
 We make a conservative estimate of this baseline by (1) taking the luminosities of all the 
 galaxies in the P3$^\prime$ region (see \textsection\ref{section_luminosity} for details),
 (2) calculating the \emph{dark matter} mass assuming a fiducial $M/L$ of 300 $M_{\sun}/L_{B\sun}$
 (higher than the M07 value of 232, and so more conservative),
 and (3) adding to this dark matter mass the maximum gas mass along the column.
For the X-ray gas mass, we adopt the upper limit $M_{gas}=0.85\times10^{13}M_{\sun}$.
The resulting mean convergence (adding both dark matter and gas masses)  within the $r<150$ 
kpc aperture is $\mytilde0.04$.
Because the FIATMAP convergence maps are subject to mass-sheet degeneracy, we
rescale the maps in such a way that the substructure masses agree with
those derived from aperture mass statistics.
The significance is computed by first subtracting the baseline value from the rescaled convergence
within the $r=150$kpc aperture and then dividing the result by the rms obtained from the 1000 runs.
The distribution in the significance of the P3$^{\prime}$ is shown in Figure~\ref{fig_sig_darkpeak}.
The mean of the distribution is $\mytilde6.6\sigma$, and the low-end tail is $>4\sigma$.
This mean value $\mytilde6.6\sigma$ is similar to the significance estimate based on our aperture 
mass densitometry (\textsection\ref{section_luminosity}).
Also displayed in Figure~\ref{fig_sig_darkpeak} is the significance distribution when we
repeat the experiment with the C12 weak-lensing catalog. The dark peak is still present in C12, although
the significance in C12 is systematically lower. The reason is two-fold. C12 used a factor of two fewer 
source galaxies, and the convergence in the P3$^{\prime}$ slightly lower.
This catalog-based comparison is detailed in 
 \textsection\ref{section_comparison_with_c12}.

\begin{figure*}
\begin{center}   
\includegraphics[width=\hsize]{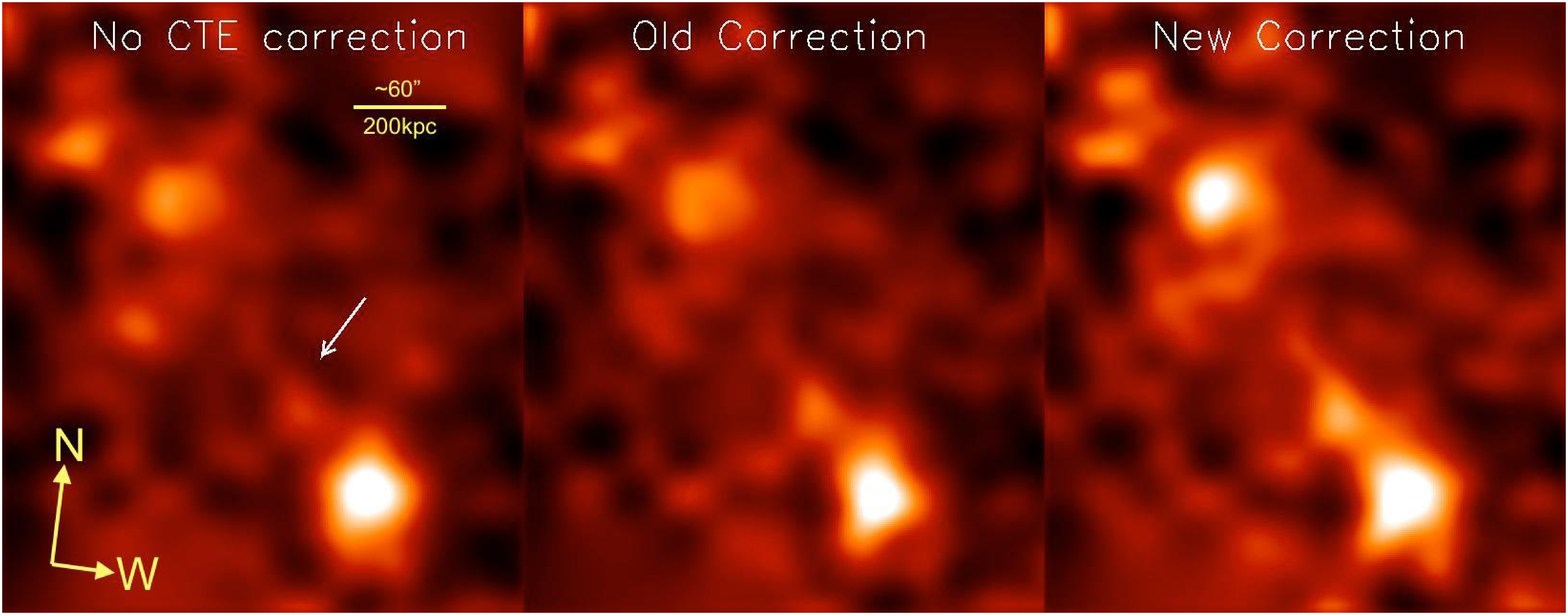}
\end{center}
\caption{Influence of CTI on the A520 mass reconstruction. We present
  mass reconstruction results based on the KS93 method for different
  CTI-correction schemes. The left, middle, and right panels
  correspond to the cases of the left, middle, and right panels in
  Figure~\ref{fig_cti_ellipticity}, respectively.  Although the CTI
  effect alone does not introduce any artificial
  substructures. Incomplete corrections will cause inaccurate
  representation of the relative strengths among different
  substructures. The arrow indicates the location which we identify as
  a new centroid of the dark core.  It is clear that the significance
  of this substructure will be underestimated if the CTI correction is
  less than optimal. We use the same intensity-to-color mapping scheme
  for the representation of the three mass maps after normalizing the
  result by the maximum convergence value of each map.
\label{fig_cti_mass_reconstruction}}
\end{figure*}

\subsection{Impact of the CTI correction model}

As discussed in \S2, CTI is an important instrumental effect and
although the Y2012 method (Ubeda \& Anderson 2012) that we use in our
analysis is not perfect, our tests on SPFs presented in
Figure~\ref{fig_cti_ellipticity} shows it performs better than the
Y2009 correction (Anderson \& Bedin 2010) down to fluxes of 300
$e^{-}$. One common misconception regarding the CTI effect on weak
lensing mass reconstruction is that it affects only the region where
the readout distance is the largest. However, this is not an accurate
statement especially in the case of two-dimensional mass
reconstructions, which are related to shear fields non-locally.
It is therefore interesting to examine whether or not
the residual CTI features affect our weak lensing analysis.

In Figure~\ref{fig_cti_mass_reconstruction} we compare the Kaiser \&
Squires (1993; hereafter KS93) mass reconstruction results when we do
not correct for CTI (left panel), use the Y2009 method (middle panel)
or the most recent approach (Y2012; right panel). Note that C12 used
the Y2009 method as well as an updated model from Massey et al. (2010)
and claimed consistent results in both cases. We assume the equality
$g=\gamma$ because an iterative nonlinear reconstruction using the
relation $g=\gamma/(1-\kappa)$ may exaggerate the difference among the
different versions. 

The overall distributions of the three convergence fields are similar
to one another without any conspicuous substructure only seen in any
particular version.  However, it is important to note that the
relative strengths of the mass peaks are significantly different. The
most critical substructure in A520 is the overdensity P3' (indicated
by the white arrow) between the two dominant mass peaks, where there
are no luminous cluster members. It is remarkable that the overdensity
in this region is strongest when we apply the best-performing Y2012
CTI correction method (right panel). The feature is still seen, but
weakest when no CTI correction is applied (left panel). The
Y2009 correction gives an intermediate significance.

As shown in Figure~\ref{fig_cti_mass_reconstruction}, the assessment
of the substructure significance relative to another is influenced by
imperfect CTI correction. Therefore, we conclude that the impact of
the fidelity of the CTI correction is non-negligible in weak lensing
analysis with the current A520 data. We revisit this issue when we
discuss our actual mass determinations in
\S\ref{section_mass_estimation}.

\subsection{Comparison with the WFPC2 analysis}

The most noteworthy difference in the weak lensing results between our
current ACS and the WFPC2 study is the large shift of the centroid of
the dark peak from P3 to P3$^{\prime}$ by $\mytilde1\arcmin$. Since
this shift is much larger than the centroid uncertainty determined
from the bootstrapping experiment (Table~1), it is difficult to
attribute the shift to mere shot noise. Here we examine whether the
shift originates from a systematic difference between the two source
catalogs of ACS and WFPC2.

\begin{figure*}
\begin{center}
\hbox{%
\includegraphics[width=0.5\hsize]{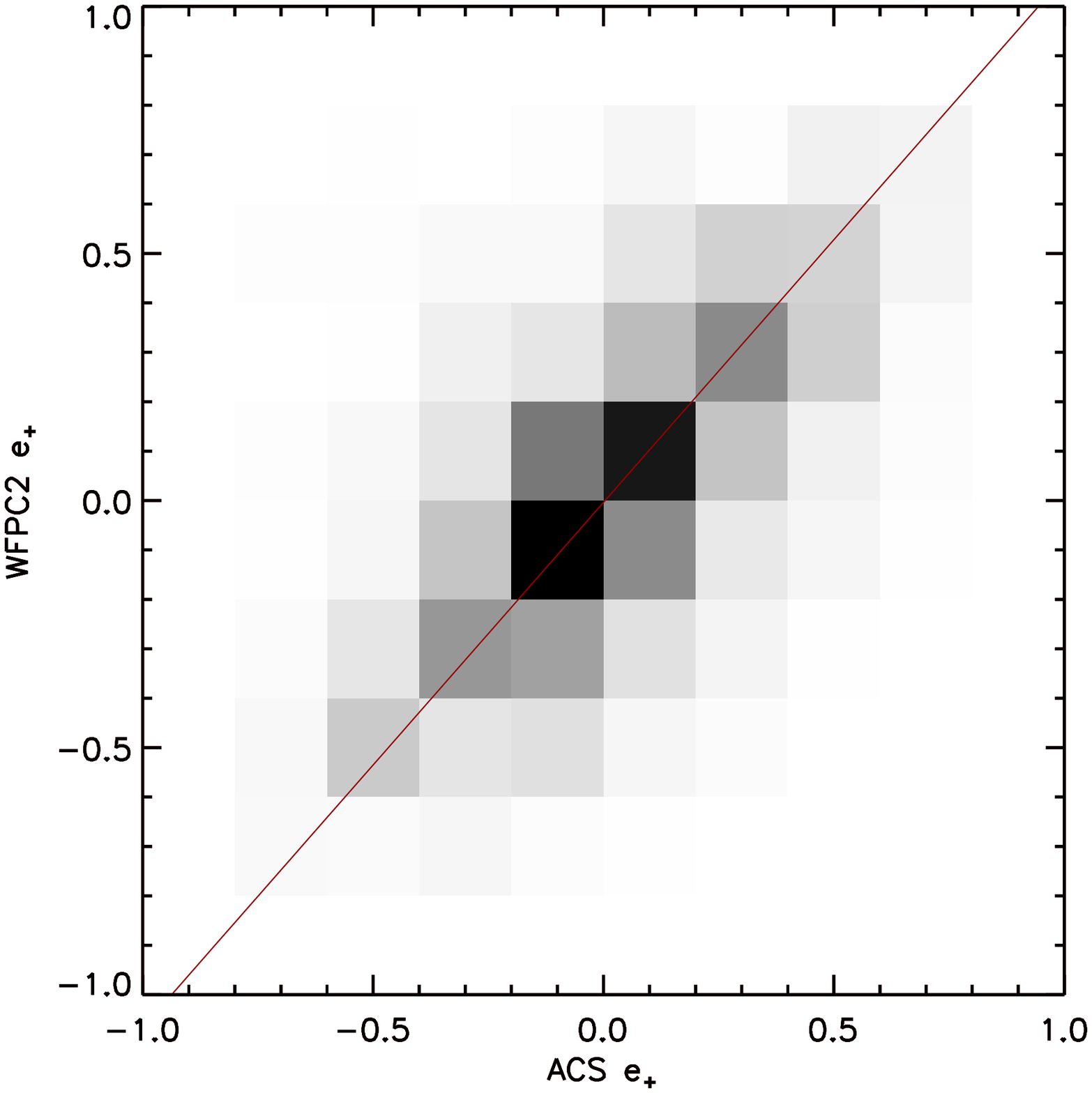}
\includegraphics[width=0.5\hsize]{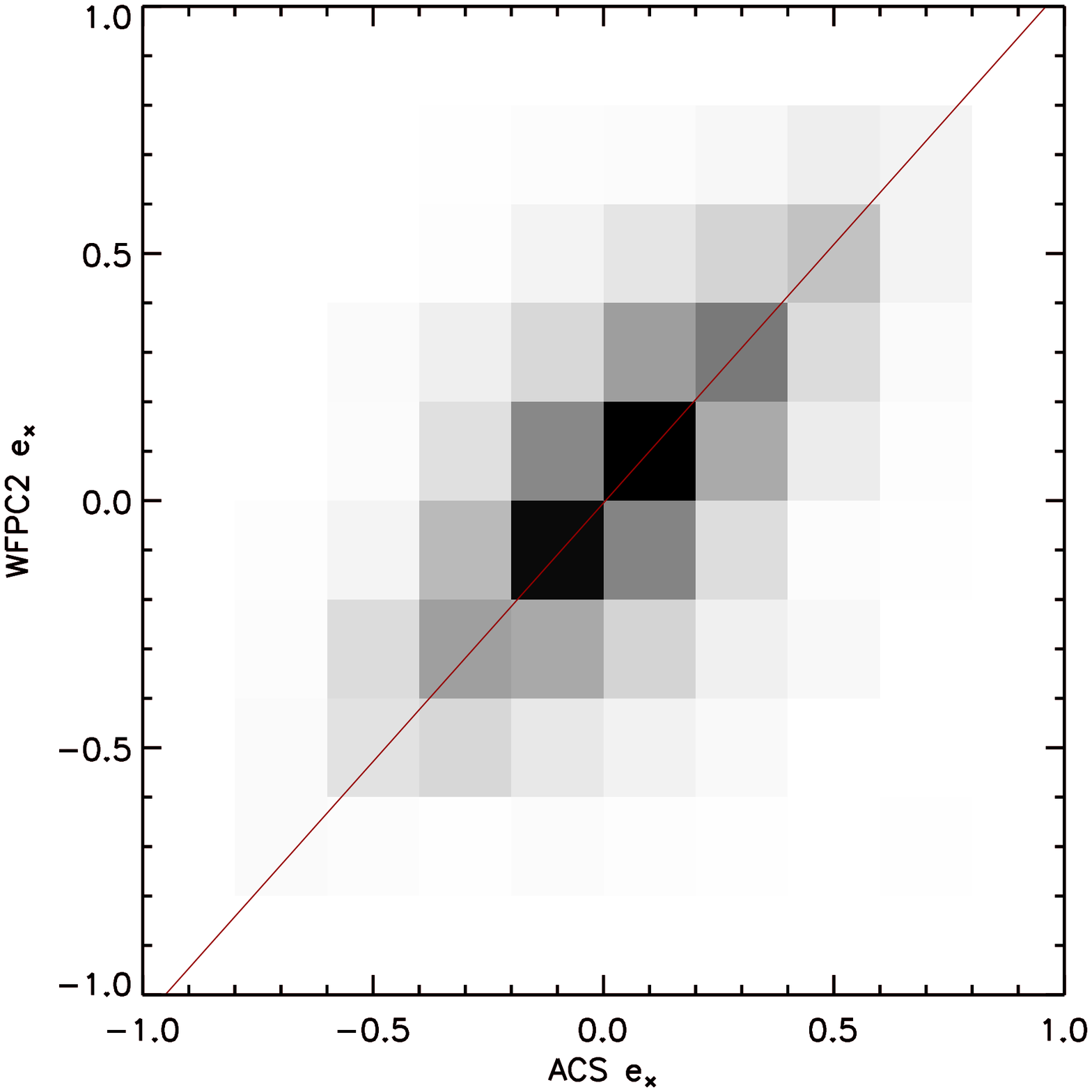}}
\caption{Ellipticity comparison between ACS and WFPC2. We compare the
  raw ellipticities that are directly obtained from elliptical
  Gaussian fitting. A total of 1764 source galaxies are common to both
  shape catalogs.  PSF and CTI effects are corrected, but no shear
  calibration is applied. The J12 ellipticities from the WFPC2 image
  agree nicely with those from the current ACS data.  The average
  slope of the two panels is $\mytilde1.05$. This small departure from
  unity disappears when we apply the due shear calibration factor to
  each dataset.
\label{fig_ellipticity_comparison_with_wfpc2}}
\end{center}
\end{figure*}

We first carry out a galaxy-by-galaxy comparison of the two catalogs.
Figure~\ref{fig_ellipticity_comparison_with_wfpc2} shows that the raw
ellipticities (prior to the application of shear calibration) between
ACS and WFPC2 agree nicely. The average slope of the two ellipticity
components is $\mytilde1.05$. This small departure from unity
disappears when we apply the appropriate shear calibration factor to each
dataset. Hence this comparison does not indicate a systematic
difference in ellipticity.

\begin{figure}
\includegraphics[width=\hsize]{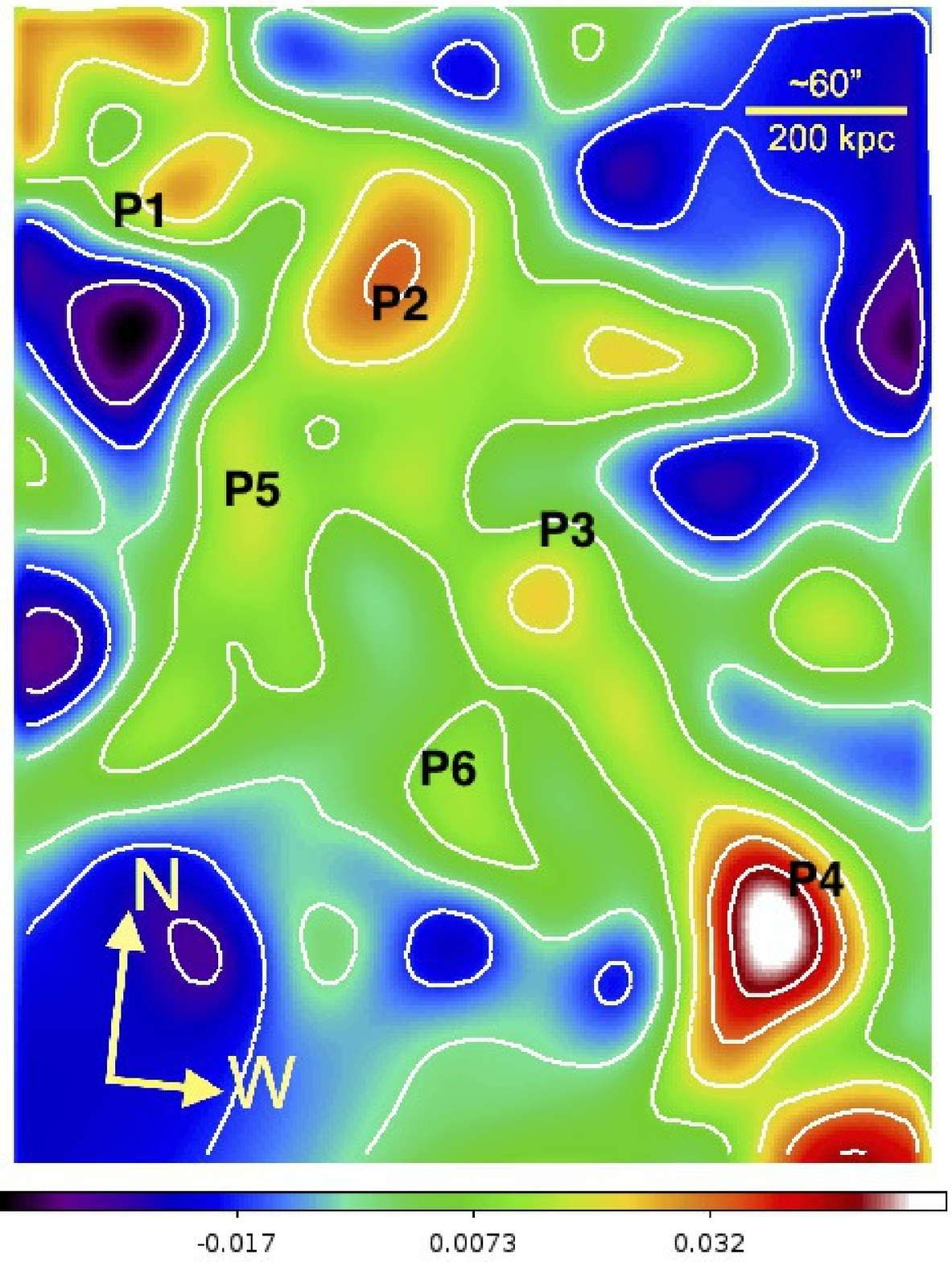}
\caption{Mass reconstruction test using the source selection in
  J12. The galaxy shapes are measured from the current ACS data.  We
  use the KS93 method to compute the convergence field by assuming
  $g=\gamma$.  Interpretation requires caution because the field
  boundary defined by the overlapping region between the ACS and WFPC2
  images is complicated. Nevertheless, we confirm that in this mass
  reconstruction the location of the dark core agrees with that in
  J12.
\label{fig_ks93_wfpc2_mass}}
\end{figure}

To examine whether any systematic shape error might be localized near
the dark peak we perform a weak lensing mass reconstruction using the
shapes of the ACS data but based on the source selection in J12.  In
our WFPC2 catalog there are more (less) source galaxies in the P3
(P3$^{\prime}$) region than in our ACS catalog. Thus, the number
density distribution of the common source galaxies does not
exactly match the source density distribution of the WFPC2
data.  We return to the issue of source density below.

Figure~\ref{fig_ks93_wfpc2_mass} shows the convergence field obtained
from this source catalog using the KS93 method. Interestingly, the
location of the dark core now coincides with that of P3 in J12 in this
mass reconstruction. This demonstrates that the shape catalogs based
on either WFPC2 or ACS data give consistent results when a similar
source selection is made.

\begin{figure*}
\begin{center}
\hbox{%
\includegraphics[width=0.45\hsize]{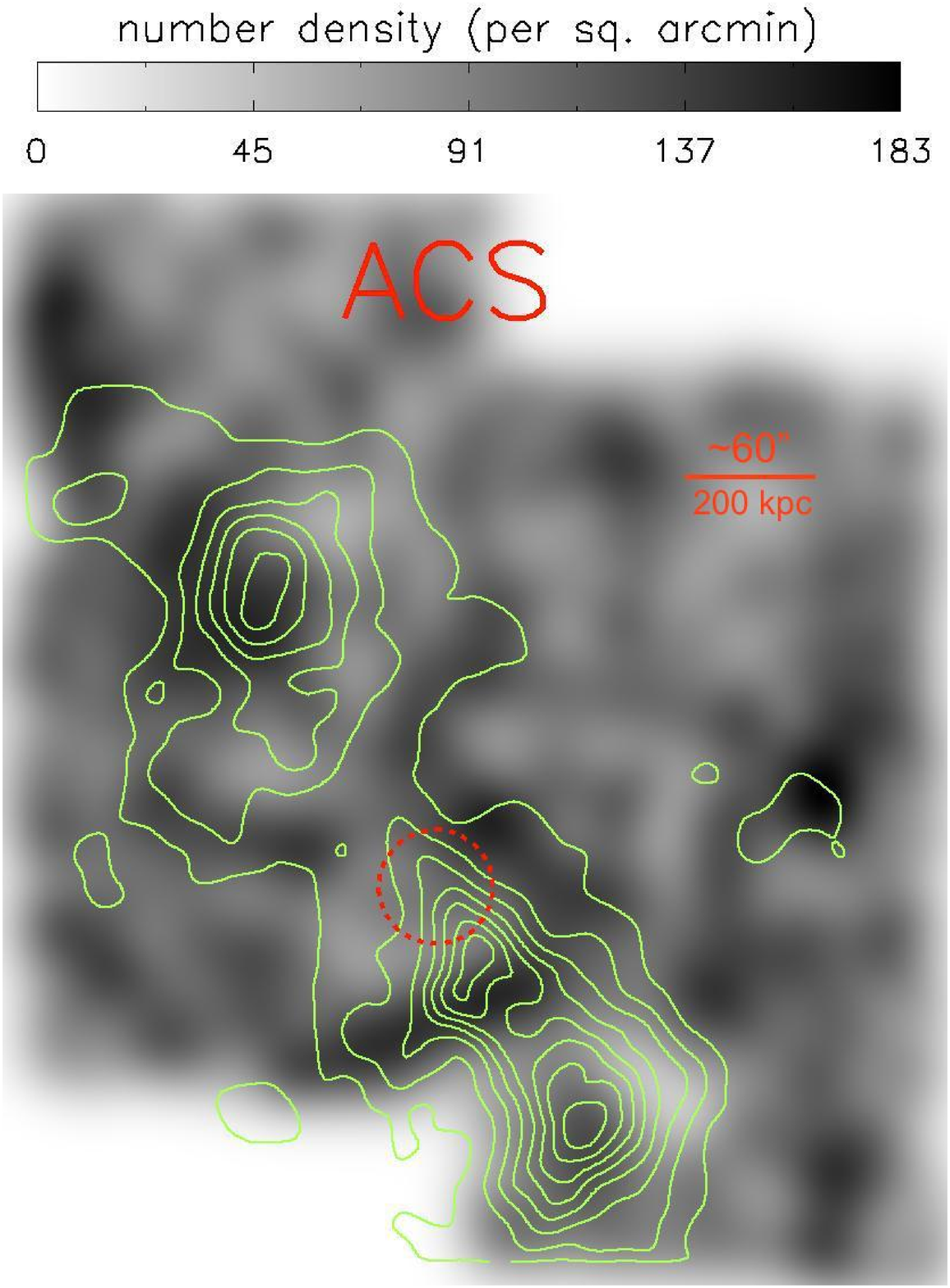}
\includegraphics[width=0.45\hsize]{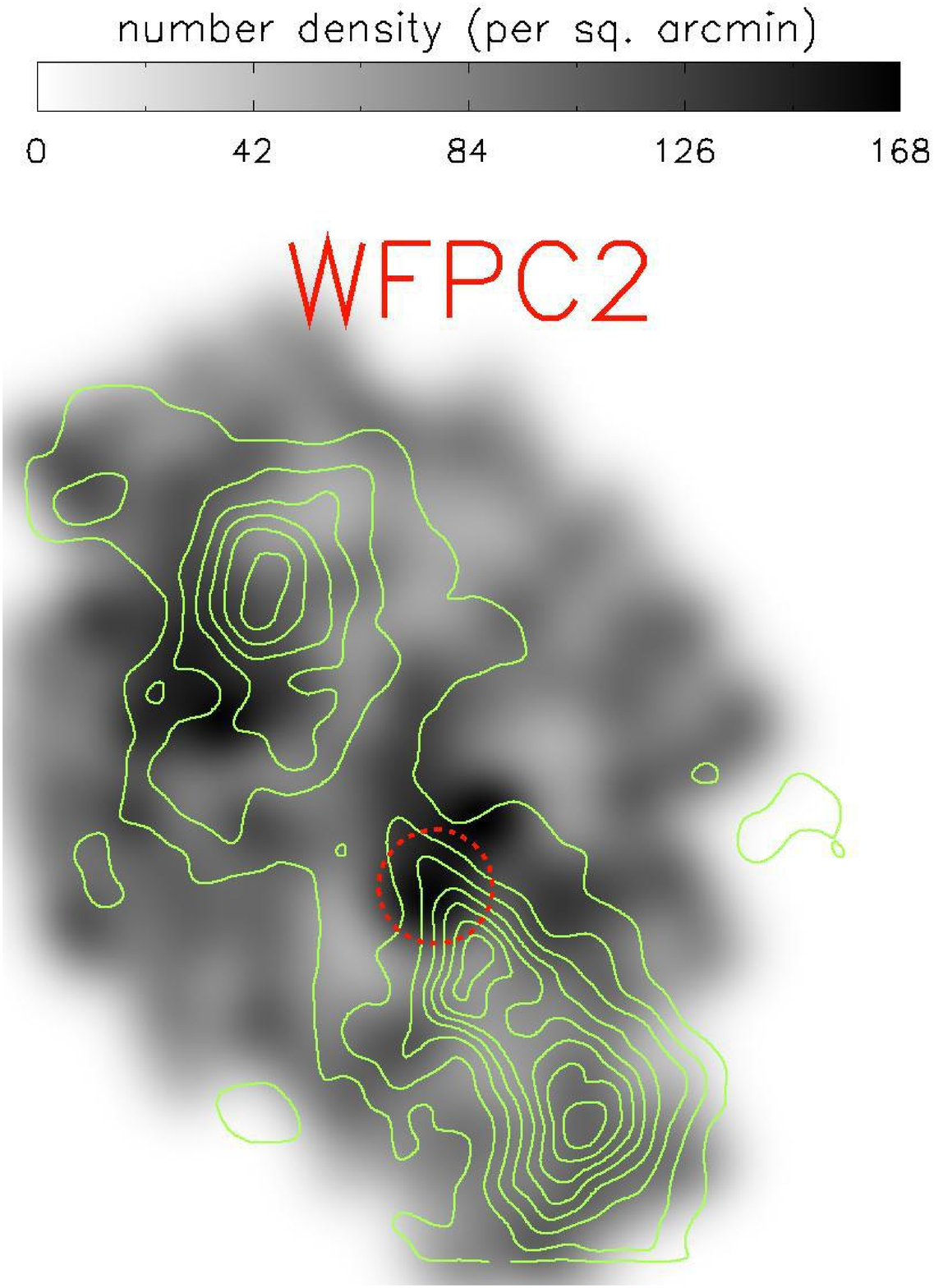}}
\caption{Difference in the source galaxy distribution between ACS
  (left) and WFPC2 (right) data. We smooth the source galaxy
  distribution using a Gaussian with FWHM=$30\arcsec$.  Overlaid are
  the convergence contours derived from the ACS data. We argue that
  the difference in the source density fluctuation might have caused
  the centroid shift of the substructure P3 in J12. One of the largest
  difference between the two source density maps is seen near P3 (red
  circle), where the WFPC2 shows a strong concentration relative to
  the neighboring region whereas this contrast is not clear in the ACS
  data.
\label{fig_source_density}}
\end{center}
\end{figure*}

\subsection{Origin of the shift in peak location}
\label{sec_shift}

While matching objects between the WFPC2 and ACS catalogs, we found
that there is a systematic difference in source galaxy $density$
distribution. Figure~\ref{fig_source_density} compares the source
galaxy distribution between the ACS and WFPC2 data. It is clear
that in the WFPC2 catalog the density in the P3 region is much higher
than the surrounding area. This is mainly because the WFPC2
observation was planned in such a way that much deeper imaging is
performed in this region (see also Figure 2 of J12).  Although we
still hold to the claim of J12 that the spatial variation of the
source galaxy number density does not cause any spurious substructure,
it appears that the centroid of the dark peak is influenced by this
large inhomogeneity of the source galaxy distribution.

We suspect that the elongation of the substructure in the dark peak
region along the merger axis also facilitates the centroid shift. This
particular shape of the substructure should make its centroid more
uncertain along the P3-P3$^{\prime}$ orientation and sensitive to the
source density fluctuation near the dark peak region. The CFHT
location of the dark peak in M07 is similar to that of J12, although
it is closer to P3$^{\prime}$ by $\mytilde20\arcsec$ than in J12.
However, given the large smoothing scale (because of the relatively
small number of source galaxies) of the CFHT mass reconstruction, we
do not consider the centroid difference between M07 and the current
study very significant (see the centroid variation in Figure 7 of M07
for different images).

Finally, we discuss the effect of cluster galaxy contamination. In
J12, the cluster member selection is based on the CFHT $g-r$ color,
mainly identifying the red-sequence galaxies of A520 except for some
blue galaxies whose spectroscopic redshifts are known. The current ACS
selection based on three filters approximately increases the number of
the cluster member candidates by $\mytilde60$\% (295 vs. 474
candidates). Most of this increase comes from increasing the number of
relatively faint blue cluster member candidates whose F435W-F606W
colors are less than $\mytilde1.3$. However, this improvement in the
cluster member removal has a negligible impact on the weak lensing
analysis for the following reasons. First, the blue cluster member
selection is still unreliable even when one uses three broadband
photometry. Second, we find that the spatial distribution of these
additional blue cluster candidates does not correlate with the
substructures. Third, most importantly, our mass reconstruction using
the source galaxy catalog where we only remove the red-sequence is
very similar to the results presented in
Figure~\ref{fig_massmap_contour}.

\begin{figure*}
\begin{center}
\hbox{%
\includegraphics[width=0.5\hsize]{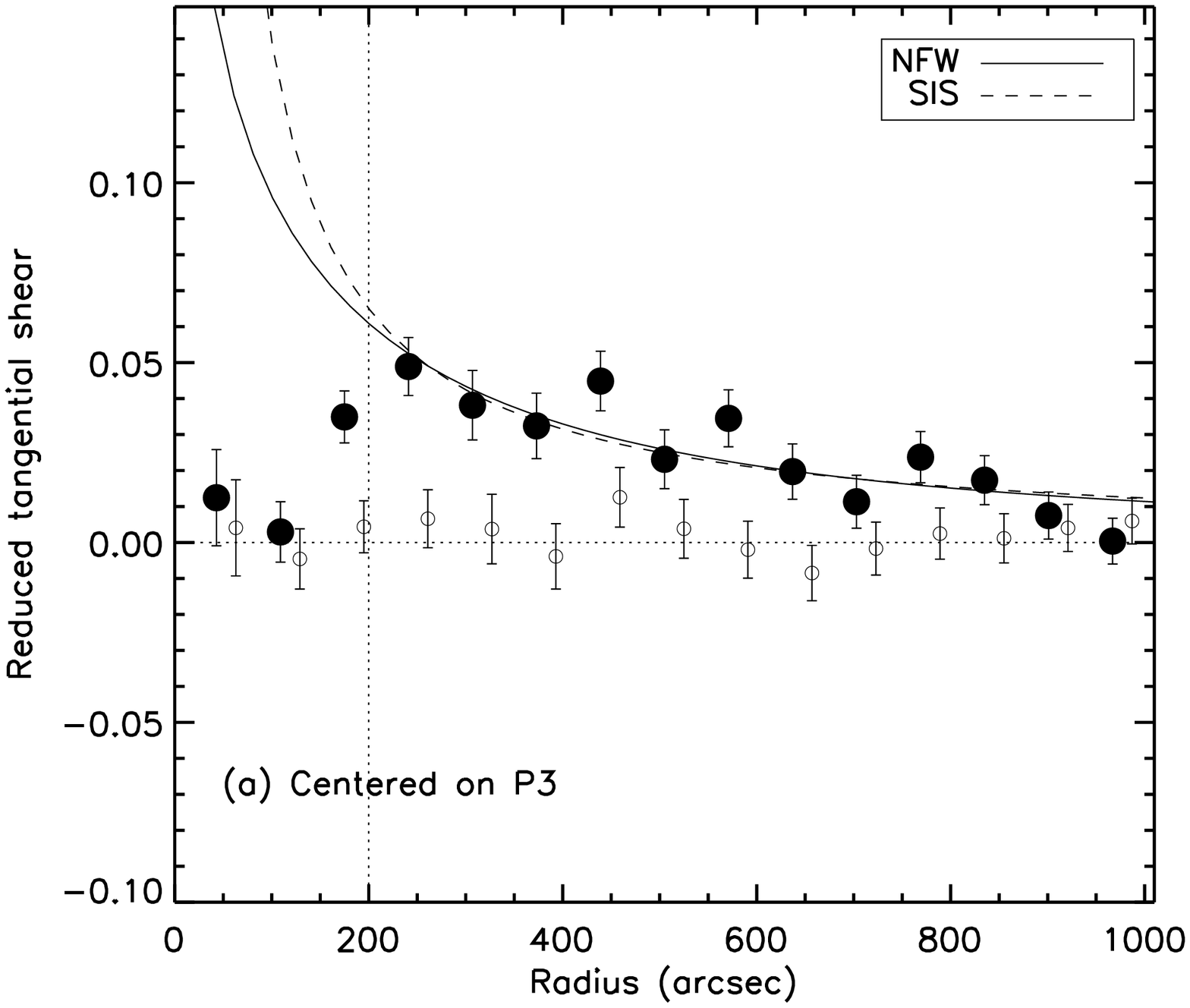}
\includegraphics[width=0.5\hsize]{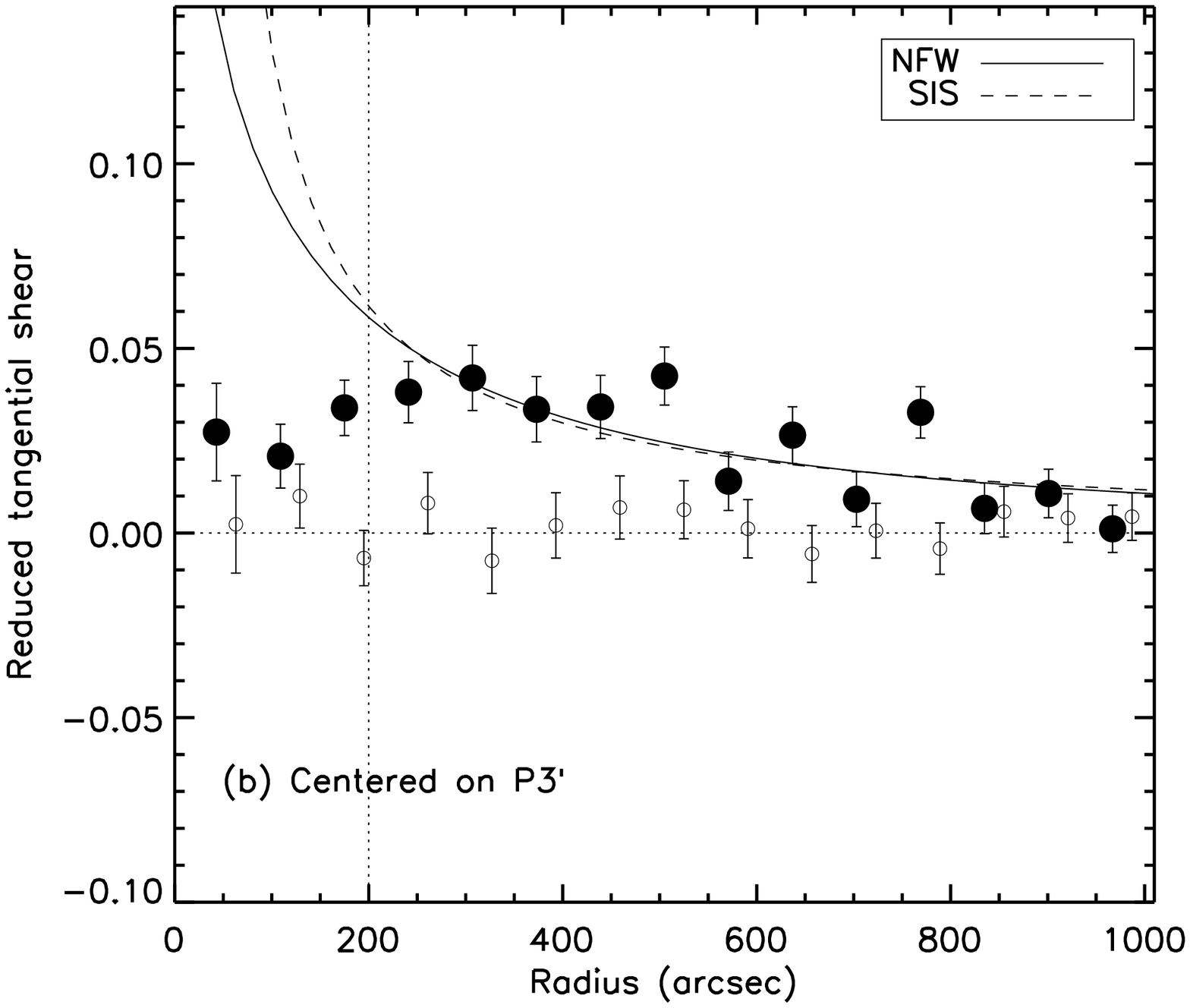}}
\end{center}
\caption{Tangential shear profile around P3 (a) and P$3^{\prime}$
  (b). We combine the shape catalogs from the ACS and CFHT images, and
  applied the due redshift scaling to the CFHT shears.  The relevant
  critical surface mass density for the plots shown here is
  $\Sigma_c=3.35\times10^3 M_{\sun} \mbox{pc}^{-2}$.  The filled
  circles represent the tangential shears around the reference
  points. The open circles represent our null text results obtained by
  rotating galaxies by 45$\degr$. The latter result (being consistent
  with zero) shows that the residual systematics in our analysis is
  negligible.  The dotted line denotes the lower limit of the radial
  bin used for fitting the two model (NFW and SIS) profiles.
\label{fig_tangential_shear}}
\end{figure*}

\section{Mass Estimates}
\label{section_mass_estimation}

It is possible to use the two-dimensional mass map to measure
substructure masses, but the results are sensitive to the algorithm
used. More importantly, on small scales the result depends on details
such as smoothing scheme, treatment of non-linearity, etc.
Also, one can consider fitting halo models (e.g., NFW) to all mass clumps 
simultaneously. This is feasible when the substructures are simple 
and sufficiently massive such as those of  the ``El Gordo" cluster (Jee et al. 2013).
We find that this simultaneous fit becomes unstable when applied to A520, which
consists of at least five mass clumps within the $r\sim0.5$~Mpc region. 

Therefore, we report our mass estimates based on the aperture mass densitometry
(Fahlman et al. 1994; Clowe et al. 2000), which provide a more direct
estimate of the projected mass by measuring the tangential shear as a
function of radius around the position of interest.

The aperture mass densitometry also allows us to take full advantage
of the large field of view provided by our ancillary Canada France
Hawaii Telescope (CFHT) data, which is important for efficiently
breaking the so-called mass-sheet degeneracy. C12 combined ACS and
Magellan data in their aperture densitometry to overcome the small
field of view of ACS. Although both this study and C12 supplement the
ACS data with ground-based images, we believe that the main
differences arise from the different treatments of the ACS data.

As is done in J12, we combine the shape catalogs from the CFHT and ACS
images by preferring ACS shapes wherever available. The difference in
the effective source plane redshift between ACS and CFHT is accounted
for by scaling up the CFHT shears by the amount required for the
difference in the redshift. We verify that the amount of correction
due the mean source redshift difference is consistent with the
difference in the amplitude of the raw tangential shear profiles in
the overlapping region. 

\noindent The tangential shear is defined as

\begin{equation}
 g_T(r)  = -  g_{+}(r) \cos 2\phi - g_{\times}(r) \sin 2\phi \label{tan_shear},
\end{equation}

\noindent where $\phi$ is the position angle of the object with respect to the
reference point. We use the symbol ``$g$" to remind readers that the
measured quantities are in fact reduced shears.  

In Figure~\ref{fig_tangential_shear}, we present the tangential shear
profiles derived from this combined shape catalog around P3 (left) and
P3$^{\prime}$ (right).  The open circles represent the results from
our ``null'' (45-deg rotation) test and show that this so-called
B-mode signal is statistically consistent with zero. The best-fit
isothermal profiles using the data at $r>200\arcsec$ (dashed) predict
a velocity dispersion of $1077\pm44~\mbox{km/s}$. The best-fit NFW
profile using the Duffy et al. (2008) mass-concentration relation
gives $M_{200}=9.6_{-1.2}^{+1.4} \times10^{14} M_{\sun}$
($r_{200}=1.91\pm0.09$ Mpc; $c=3.21\pm0.04$)\footnote{By $M_{200}$, we
  define the total within the radius for which the mean internal
  density is 200 times the critical density.}.  Although we quote
these values based on the tangential shears around P3, little
difference is observed as to the total mass of the cluster when we
select P3$^{\prime}$ as a reference point.  C12 report
$M_{200}=(9.1\pm1.9)\times10^{14} M_{\sun}$ by assuming $c=3.5$. Their
$M_{200}$ mass is consistent with ours. 

A good agreement in terms of the global mass is also seen when
we compare projected masses. C12 estimate that within $r=700$ kpc the
aperture mass is $(5.1\pm0.7)\times10^{14} M_{\sun}$, which is
close to our estimate $(5.72\pm0.36) \times10^{14} M_{\sun}$.  

The aperture mass statistics can be evaluated by performing the
following integral:

\begin{eqnarray}
\zeta _c (r_1,r_2,r_{max}) & = & \bar{\kappa}( r \leq r_1) -
\bar{\kappa}( r_2 < r \leq r_{max}) \\ &= & 2 \int_{r_1} ^{r_2} \frac{
  \left < \gamma_T \right > }{r}d r + \frac{2}{1-r_2^2/r_{max}^2}
\int_{r_2}^{r_{max}} \frac{ \left <\gamma_T \right >}{r} d r, 
\label{eqn_aperture_densitometry}
\end{eqnarray}

\noindent where $\langle \gamma_T \rangle$ is the azimuthally averaged
tangential shear, $r_1$ is the aperture radius, and $r_2$ and
$r_{max}$ are the inner- and the outer radii of the annulus.  It is
important to iteratively update $\gamma$ using $\gamma=(1-\kappa)g$
where $\kappa$ is non-negligible.  Because $\zeta_c(r_1,r_2,r_{max})$
provides a density contrast of the region inside $r<r_1$ with respect
to the control annulus $(r_2,r_{max})$, one desires to choose $r_2$
and $r_{max}$ to be large so that the mean density in the control
annulus becomes small and mostly limited by the large scale structure
(i.e., cosmic shear) of the field.  In this paper, we choose the
annulus defined by $r_2=600\arcsec$ and $r_{max}=800\arcsec$.  When P3
is selected as the reference, the mean density of this region is
estimated to be $\bar{\kappa}=0.018$ ($\bar{\kappa}=0.009$) according
to the SIS (NFW) profile fitting result
(Figure~\ref{fig_tangential_shear}).  This factor of two difference in
the density estimates of the annulus causes only a $\mytilde5$\%
difference in the substructure masses within the $r=150$ kpc
radius. In this paper, we adopt the SIS fitting values for consistency
with J12.

As for error propagation in aperture mass densitometry,
we perform 1000 Monte Carlo simulations by randomizing the
tangential shear profiles. 
Neither the difference in the
background density estimation nor the effect of the cosmic shear
(Hoekstra 2001, 2003) is included in our error propagation. We note
that the latter is not important on small scales.

Table~1 lists the substructure masses obtained from the current
aperture mass densitometry.  We leave out the substructure P6 because
its $r=150$kpc circle substantially overlaps with that for
P3$^{\prime}$. We do list the results for P3 to enable a comparison
with the results from J12 and C12.

The aperture mass centered on P3$^{\prime}$ is $(3.94\pm0.30)\times
10^{13} M_{\sun}$, whereas for P3 we find $(3.35\pm0.34)\times 10^{13}
M_{\sun}$. The latter value is lower than that of J12 by
$\mytilde16$\%, but the error bars of the two results overlap. C12
quotes an even lower value $(2.84\pm0.64)\times 10^{13} M_{\sun}$ from
their aperture mass densitometry. This estimate is statistically
consistent with our current result, mainly because of their larger
errors. When we use the ACS shape catalog matched to the WFPC2 observations,
we obtain $(3.66\pm0.35)\times10^{13} M_{\sun}$, in better agreement with
the J12 value.

As both this study and C12 analyze the same
ACS data, perhaps this 16\% discrepancy in the central value is an
indicator of a systematic difference between the two studies.  The C12
mass of P3 is in slight tension with the J12 and M07 values at the
$\mytilde$2$\sigma$ level. A similar level of variations exists for
other substructures as well.  For example, the current aperture mass
of P4 is $(4.23\pm0.30)\times 10^{13} M_{\sun}$, whose central value is
about 16\% higher than the J12 result, although again the two results
are statistically consistent.  C12 estimated the mass of P4 to be
$5.59\pm0.68$. This result is $\mytilde32$\% higher than
the result estimated in this study, and the discrepancy is
larger than in the case of the P3 comparison.

The ratio of the aperture mass of P3$^{\prime}$ to that of P3 in our
ACS analysis is $\sim 1.17$, which may appear discordant with the
visual impression of a larger difference that one receives from the
mass reconstruction (Figure~\ref{fig_massmap_contour}).  We attribute
this difference to the different range of tangential shears affecting
each result: the aperture mass densitometry uses only the information
outside the aperture radius ($r>r_1$) whereas the mass reconstruction
is influenced by the shear signal inside the aperture, as well as the
information outside the aperture. Therefore, the factor of two higher
amplitude of the tangential shear (Figure~\ref{fig_tangential_shear})
at the inner most bin, which is not included in the aperture mass
statistics, is responsible for the larger contrast between P3 and
P3$^{\prime}$ in our ACS mass reconstruction.

In \textsection\ref{section_cti_impacts} we demonstrate that the
fidelity of the CTI correction non-negligibly affects the mass
reconstruction results.  The substructure around the dark peak region
becomes strongest when the latest Y2012 model (Ubeda and Anderson
2012) is applied. The feature becomes weaker when the Y2009 CTI
correction method (Anderson \& Bedin 2010) is used, and it becomes
weakest when we do not apply any CTI correction. We observe a
consistent trend in aperture mass densitometry. With the Y2009 CTI
correction, we obtain $M_{ap} (r<150~\mbox{kpc})=(3.09\pm0.33) \times
10^{13} M_{\sun}$ for P3, $\mytilde8$\% lower than the above
$(3.35\pm0.34)\times 10^{13} M_{\sun}$.  When no CTI correction is
applied, the resulting aperture mass becomes $M_{ap}
(r<150~\mbox{kpc})=(2.87\pm0.43) \times 10^{13} M_{\sun}$,
$\mytilde14$\% lower than the result that we obtain with the latest
CTI model. On the other hand, we find that the aperture mass of P4
increases by $\mytilde10$\%, when no CTI correction is applied.

\begin{deluxetable*}{lcccccc}
\tabletypesize{\scriptsize}
\tablecaption{Optical Luminosity of Substructure ($r<150$~kpc)}
\tablenum{2}
\tablehead{\colhead{Substructure} & \colhead{$L_{B}$} & \colhead{$L_{R}$} & \colhead{$L_{F814W}$}    & \colhead{$M/L_{B}$}    & \colhead{$M/L_{R}$}   & \colhead{$M/L_{F814W}$}  \\  
                    \colhead{}   &  \colhead{($h_{70}^{-2} 10^{11} L_{B \sun}$)}  &   \colhead{($h_{70}^{-2} 10^{11} L_{R \sun}$)}   & \colhead{($h_{70}^{-2} 10^{11} L_{F814W \sun}$)} &                   
                    \colhead{($h_{70} M_{\sun}/L_{B \sun}$)}  &        \colhead{($h_{70} M_{\sun}/L_{R \sun}$)}    &    \colhead{($h_{70} M_{\sun}/L_{F814W \sun}$)}           \\}
\tablewidth{0pt}
\startdata
P1  & 1.35 & 1.52 & 2.18 &                                  $139\pm32$ &    $123\pm28 $   & $86\pm20$ \\
P2  &  3.07 & 3.31 & 4.72 &                                $119\pm 9$ &     $110\pm8$      & $77\pm6$ \\
P3  & 0.80 & 0.98  & 1.40 &        			$326\pm43$  &   $266\pm34$    & $186\pm24$ \\
P3' &  0.32 & 0.38 & 0.54  & 				$966\pm97$ & $813\pm78$ &  $572\pm55$ \\
P4  &  3.36 & 3.83  & 4.13 &                             	$116\pm9$      &  $102\pm7$  & $94\pm7$  \\
P5  &  2.04 & 2.53 & 2.96   &                                 $133\pm19$ & $107\pm15 $  &  $92 \pm13$ 
\enddata
\tablecomments{We subtract the gas mass (the upper limit in Table 1) in the estimation of the M/L values.
For the estimation of $L_{F814W}$, no color transformation is performed in order to ease the comparison with the C12 results.
}
\end{deluxetable*}

\begin{figure}
\begin{center}
\includegraphics[width=\hsize]{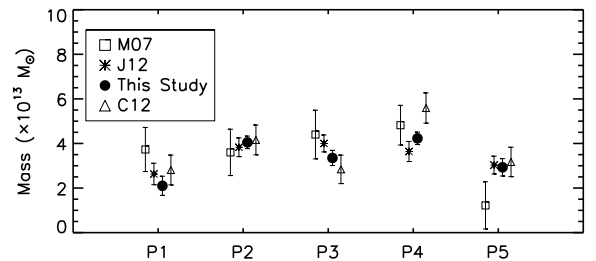}
\end{center}
\caption{Mass comparison among four different studies. We compare
  aperture masses from M07, J12, C12, and this study. The two largest
  differences between this study and C12 are found for the mass
  estimates of P3 and P4. C12 give a lower value for P3 and a higher
  value for P4. We are able to reproduce this trend when we repeat our
  weak lensing analysis without performing any CTI correction (see
  also Figure~\ref{fig_cti_mass_reconstruction}).  Not compared in
  this plot is the substructure mass of P3$^{\prime}$, which is not
  identified by C12.
\label{fig_mass_comparison}}
\end{figure}

We compare the substructure masses in Figure~\ref{fig_mass_comparison}
with those obtained by J12 and C12. Our ACS weak lensing analysis of
A520 provides results that are generally consistent with those from
our previous WFPC2 study (J12). The central values of the masses of
P1, P3, and P5 are lower in our ACS study by $\mytilde20$\%,
$\mytilde16$\%, and $\mytilde3$\%, respectively while those of P2 and
P4 are lower in our previous WFPC2 study by $\mytilde5$\% and
$\mytilde14$\%, respectively.  Considering the error bars attached to
these values, none of the differences causes a serious
tension. Nevertheless, it is worth noting that our current ACS results
give a more uniform distribution of the M/L values (see below) for the
substructures except for the dark mass region.

\begin{figure}
\begin{center}
\includegraphics[width=\hsize]{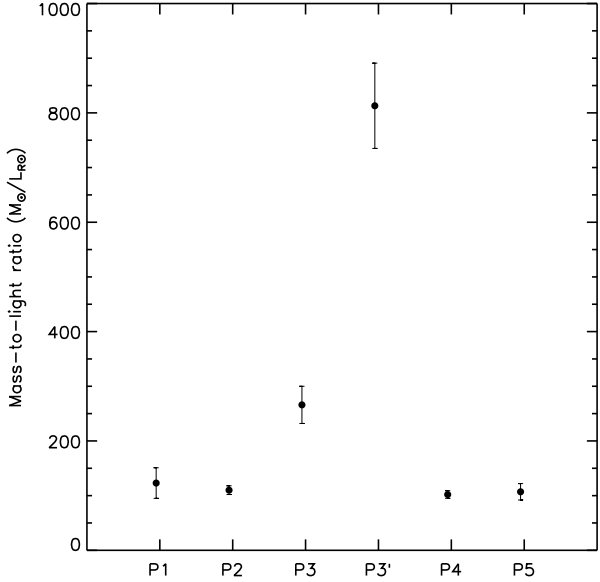}
\end{center}
\caption{Mass-to-light ratio comparison among the different
substructures in A520. The M/L values of the ``normal" peaks (P1, P2, P4, and P5) are
consistent with one another with small scatters around their mean  $\mytilde112M_{\sun}/L_{R\sun}$.
However, the dark peaks (P3 and P3$^{\prime}$) have considerably higher M/L values.
\label{fig_ml_comparison}}
\end{figure}

\subsection{Luminosity and $M/L$ Estimation} \label{section_luminosity}

M07 and J12 estimated the rest-frame $B$-band luminosity by selecting
cluster galaxies using the CFHT $g-r$ color in conjunction with the
spectroscopic catalogs.  In this paper, we update the luminosities of
A520 using the ACS data. The availability of three filters and the
improved photometry thanks to high-resolution imaging are expected to
improve the accuracy in the A520 luminosity estimation. We define the
cluster galaxies as the objects whose F606W-F814W and F435W-F606W
colors are consistent with those of the spectroscopic members. 

For easy comparison, we adopt the same quadrilateral boundary shown in
Figure 1 of C12.  We discard the object if the spectroscopic
redshift is known and is not within the cluster redshift range. Also,
stars are identified by comparing the light profile with that of the
model PSF.  The F814W filter is close to the rest frame $R$ filter at
$z=0.2$, and we establish the photometric transformation by performing
synthetic photometry using the spectral energy distribution (SED) templates 
of Kinney et al. (1996)\footnote{We use the SEDs of the elliptical, S0, Sa, Sb, SB1, and SB2 galaxies from
the Kinney-Calzetti spectral atlas.}
and the filter throughput curves of ACS and
Johnson R; we verify that the elliptical template of Kinney et al. (1996) yields F606W-F814W$\simeq0.78$ and 
F435W-F606W$\simeq1.73$, consistent with the observed values of the A520 spectroscopic members (see Figure 1 of C12).

The linear best-fit result is:
\begin{equation}
R_{\rm rest} = F814W - 0.083 (F606W-F814W) + 0.42 - DM,
\end{equation}
\noindent where $DM$ is the distance modulus for the cluster
redshift. We summarize the rest-frame luminosity and the resulting M/L
value of the cluster substructure in Table~2.  We update the
$B_{\rm rest}$ luminosity of J12 based on the current new cluster member
selection, and the results are also displayed in Table~2. 

C12 did not perform any $k$-correction and converted
their observed F814W magnitude into the rest-frame luminosity by
simply applying the distance modulus (D. Clowe, in private
communication). We refer to this luminosity as $L_{F814W}$ and list
our estimate also in Table~2. Our estimate of the luminosity for P3
when we ignore the color-dependence of the $k$-correction is
$\mytilde1.40 h_{70}^{-2} 10^{11} L_{F814W \sun}$, in agreement with
the C12 estimate. The luminosities for the other peaks agree to $\sim
15\%$, with the exception of P4, where C12 find a $33\%$ lower
value. However, the location of P4 in C12 is offset from the peak in
the luminosity distribution, whereas the position in our mass
reconstruction coincides better with the light. C12 mention that the
luminosity increase by $\sim 17\%$ when the center on the peak of the
light distribution, in line with the variation we see for the other
peaks. Furthermore, to facilitate the comparison with the mass
reconstruction, C12 measure luminosities from the smoothed light map.
Although this does not bias their mass-to-light ratios, it does reduce
the luminosity measured within a fixed aperture by $\sim 5-10\%$.

The luminosities used by C12 are nearly a factor of two higher than
those of J12, but our comparison indicates that this difference is
mostly due to the fact that they list results for a redder band and
the lack of the $k$-correction. The Kinney et al. (1996)
SED of the elliptical galaxy gives more flux in $R$ than $B$ by
$\mytilde35\%$ when normalized with the SED of the Sun. In fact, as
discussed above, we can reproduce the C12 results. 

C12 argue that their selection criteria (we adopt the same criteria
also in our current study) are more inclusive of blue cluster galaxy
candidates than those of J12. The comparison of $L_{B\sun}$ between
this study and J12 shows that the difference is small, and our our
updated rest-frame $B$-band luminosity for P3 is only 18\% higher
than the J12 value. This is in part because the $g-r$ color selection
window in J12 was broad enough to include most of the bright galaxies
selected in the current study.  According to the current selection,
the $B$-band luminosities of P3 and P4 increase by $\mytilde18$\% and
$\mytilde14$\%, respectively.  On the other hand, the $B$-band
luminosities of P1, P2, and P5 are reduced by $\mytilde12$\%,
$\mytilde17\%$, and $\mytilde4$\%, respectively.

For the evaluation of the mass-to-light ratios, listed in Table~2, we
subtracted the upper limit of the gas mass. For the luminous
substructures P1, P2, P4, and P5 we find M/L values that are
consistent with one another, yielding a mean M/L in the $R$-band of
$\mytilde114 M_{\sun}/L_{R\sun}$. For the $B$-band we find a value of
$\mytilde 131 M_{\sun}/L_{B\sun}$.

For sample of 4 clusters Hoekstra et al. (2002) found an average
mass-to-light ratio of $279\pm21 M_{\sun}/L_{B\sun}$ (evaluated at
z=0.2, assuming passive evolution). Sheldon et al. (2009) examined a
large sample of clusters observed in the Sloan Digital Sky Survey.
Figure~8 in Sheldon et al. (2009) shows that on small scales the
central galaxies dominate the light, resulting the mass-to-light to
increase with radius, before leveling off beyond $\sim 1$Mpc.
Assuming passive evolution, and taking their highest richness bin, for
which the mean mass resembles that of A520 well, their results imply a
value of $424\pm29 M_{\sun}/L_{i\sun}$ for the cluster. The average
asymptotic value for the range in richness is $293\pm8
M_{\sun}/L_{i\sun}$. 

Sheldon et al. (2009) examine the M/L as a function of distance from
the BCG and their results suggest that within the inner 150 kpc, the
mean M/L is about 40-50\% of the asymptotic value. The values listed
in Table~2 for the luminous substructures are in good agreement with
this finding if we consider the M/L values from Hoekstra et
al. (2002).  On the other hand, the M/L values of P3 and P3$^{\prime}$
are much higher.  Compared to
the global value $232\pm25M_{\sun}/L_{B\sun}$ in M07\footnote{We cannot estimate the 
global M/L with the current ACS
data because of the field limit.},
the M/L in P3 is
higher by $2.4\sigma$, but value for P3$^\prime$ is more than
$7\sigma$ higher. Our ACS weak lensing analysis therefore supports the claim of
M07 and J12 for the presence of substantial dark mass in this
region. In Figure~\ref{fig_ml_comparison}, we compare the M/L values of the substructures in A520.

\section{Detailed Comparison with C12} \label{section_comparison_with_c12}

We have compared the results from our weak lensing analysis of ACS
data to the results from J12 which was based on WFPC2 observations. In
this section we present a more detailed comparison with the results
presented in C12.  We start by noting that the overall large-scale
distribution of the C12 mass map is similar to our ACS result.
However, there are a few differences worthy of further discussion.

The C12 mass map shows some indication of overdensity at the location
of P3$^{\prime}$, but it appears more as an extension of P4, rather
than a definite peak as seen in our mass reconstruction
(Figure~\ref{fig_massmap_contour}). However, our tests of the 
CTI correction suggest that this may be the main cause of this difference.
C12 also noticed a substructure about $2\arcmin$ north of P4
and labeled it as "Peak 7" (Figure 2 of C12). However, this mass peak
does not appear in our mass reconstruction, although there is a
weak indication of an overdensity around the location in our mass map.

The M/L values of the substructures are slightly different. The
M/L value of the P3 region ($150\pm44 M_{\sun}/L_{F814W\sun}$) from
C12 is lower than the current value ($200\pm24
M_{\sun}/L_{F814W\sun}$) by $\mytilde20$\%, although the two error
bars marginally overlap.  This happens because 1) their aperture mass
of P3 is lower than ours by $\mytilde$15\% and 2) their gas mass
estimate $0.69\times 10^{13} M_{\sun}$ is higher than our value
$0.52\times 10^{13} M_{\sun}$ by $\mytilde$33\%.  The error bars of
the aperture mass of P3 between C12 and this paper also
overlap. Nevertheless, because these two results are derived from the
same ACS data, it is difficult to attribute this 15\% difference
solely to statistical noise arising from different source galaxy
selection.  C12 argue that the main difference in the M/L values
between C12 and J12 comes from the discrepancy in the luminosity
estimates, claiming that their luminosity estimate around P3 is a
factor of two higher than that of J12 because their cluster member
selection based on three ACS filters includes more blue cluster
members that J12 might have missed. However, as already mentioned in
\textsection\ref{section_luminosity}, we find that the factor-of-two
discrepancy in luminosity arises mainly from the difference in the
passband and the omission of the due $k$-correction.  Using the C12
selection criteria, we estimate $0.80\times10^{11} L_{B\sun}$ in the
rest-frame $B$, which is only $\mytilde18$\% higher than the estimate
of J12.  Ignoring the $k$-correction, we obtain $1.40\times10^{11}
L_{F814W\sun}$ in the rest-frame F814W, which is in good agreement
with the estimate of C12.

C12 presented their bootstrap resampling experiments and claimed that
any substructure resembling the dark peak only happens in
$\mytilde$2\% of the total realizations. C12 speculated on the
possibility that some chance alignment of the sources might have led
to the detection in previous studies. However, the source number
density in J12 data is significantly higher than that of M07 and Okabe
\& Umetsu (2008). Furthermore, our ACS study is based on a higher
source density compared to C12 and we confirm the presence of an
overdensity. A caveat is that C12 focused on the location of the dark
peak defined in J12, which is $\mytilde1\arcmin$ offset from the
current centroid. C12 might have obtained different results if they
had examined the region near P3$^\prime$. As stated in
\textsection\ref{section_mass_reconstruction}, our bootstrapping
test shows that the dark peak P4$^{\prime}$ appears $\mytilde99$\%
of the random realizations at the $>4\sigma$ significance.

We note that similar levels of M/L and mass discrepancy are present in
other substructures as well. For instance, C12 quotes an aperture mass
of $(5.59\pm0.68)\times10^{13}~M_{\sun}$ for P4 whereas we estimate
$(4.23\pm0.28)\times10^{13}~M_{\sun}$. This $32\%$ difference for P4
is in fact larger than the contrast in P3.
Figure~\ref{fig_mass_comparison} shows the comparison for the rest of
the substructures.

The error bars of C12 are on average a factor of two larger than
ours, and here we provide detailed analysis of the discrepancy.
C12 present analytic expression for estimating the $\zeta$ statistic
as follows:
\begin{equation}
\sigma_{\zeta}^2= \left ( \frac{ 2  (d\ln{r})~ \sigma_{SN} }{1-r_1^2/r_{max}^2} \right )^2 \Sigma n_{bin}^{-1},
\end{equation}
\noindent
where $n_{bin}$ is the effective number of sources per logarithmic bin. The summation is carried
out over these logarithmic bins.
Note that we correct for
the typographical error in C12, where the exponent of the $(1-r_1^2/r_{max}^2)$ term should have been two (as above)
not one. The equation is an approximation because 1) the integral in the
$\zeta$ statistic is treated as summation, 2) the aperture mass is estimated
using $\zeta_c$ rather than $\zeta$, and 3) the nonlinearity $g=\gamma/(1-\kappa)$ is
ignored. We compare the results of this analytic error propagation
with those obtained from our direct Monte-Carlo analysis and find that
the approximation overestimates the errors by $70\sim90$\% for the $r=150$~kpc aperture mass
given the same source density.
The remaining discrepancy comes from the difference in
the number density of source galaxies. The number density in C12 is 56 per sq. arcmin whereas it is 109 per sq. arcmin
in this study. De-weighting low S/N
galaxies being considered, the rms shear of C12 is $\mytilde30$\% higher than ours (\textsection\ref{section_source_selection}).

We find that the small source galaxy density of
C12 is rather surprising because the typical source density in our
previous weak lensing studies with HST/ACS images comfortably exceeds
$\mytilde100$ per sq. arcmin whenever the number of orbits per
pointing is two or higher as in the current A520 data (Jee et
al. 2005a; 2005b; 2006; 2007; 2011; Dawson et al. 2012). Also, even
for ACS images with the depth of a single orbit, our typical number
density of usable galaxies is above $\mytilde70$ per sq. arcmin (Jee
et al. 2011). The source number density also depends on
shape measurement and image reduction method. It is possible that the C12 implementation
of KSB together with the use of the square drizzling kernel
gives a smaller number of usable galaxies.  Nevertheless,
Schrabback et al. (2010), who also use a KSB technique, still quote
$\mytilde$76 galaxies per sq. arcmin from their analysis of the COSMOS
data, where the mean number of orbits per pointing is about one.

Finally, we carry out a catalog-level comparison between this study and C12.
C12 kindly agreed to exchange each team's shape catalogs to enable
a detailed comparison.
The total number of sources in C12 is 2,507 whereas it is 4,788 in
this study. By looking for pairs within the distance of $0\farcs5$, we identify 
2,148 common objects between the two source catalogs. 

\begin{figure*}
\begin{center}
\includegraphics[width=8cm]{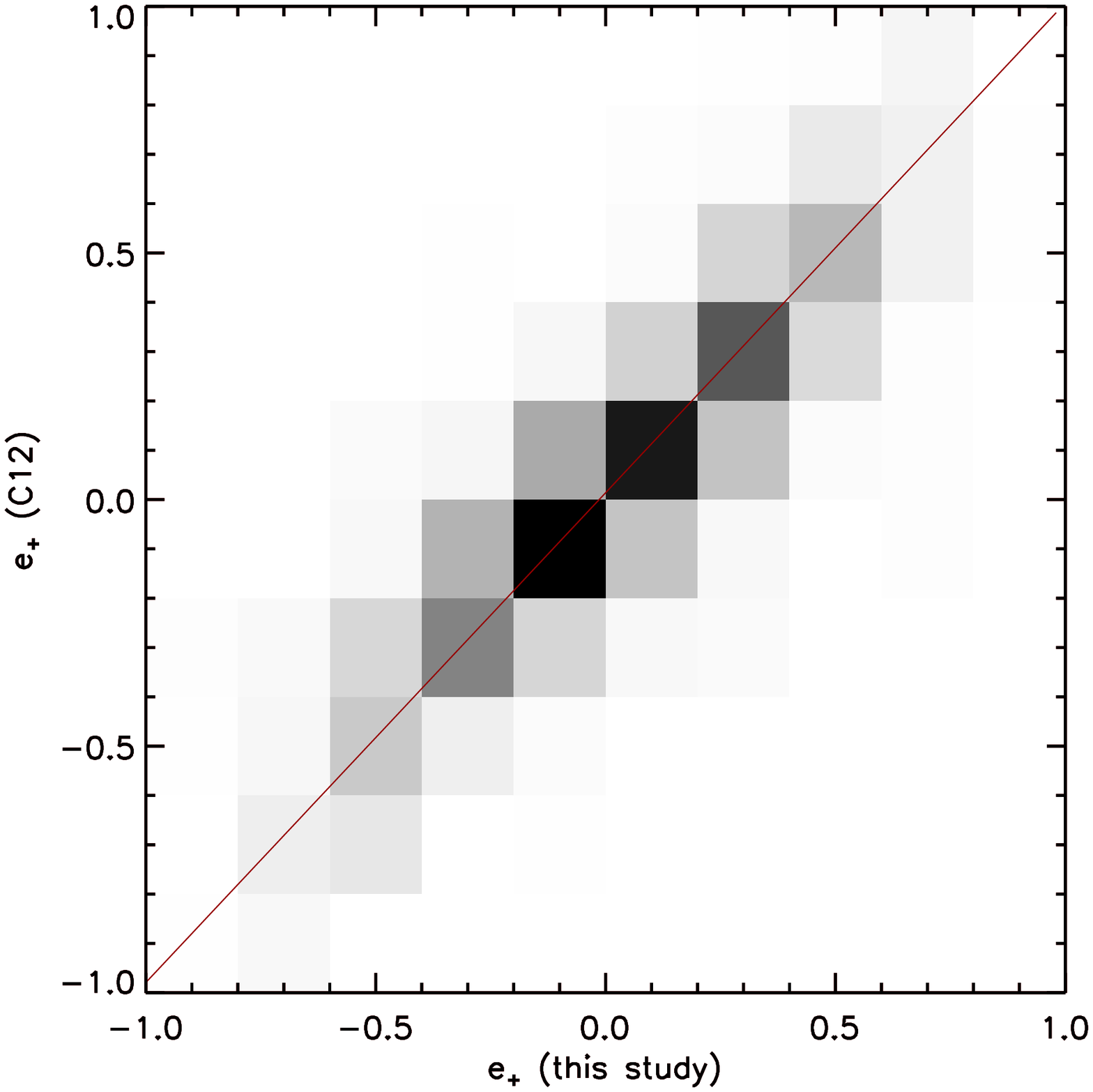}
\includegraphics[width=8cm]{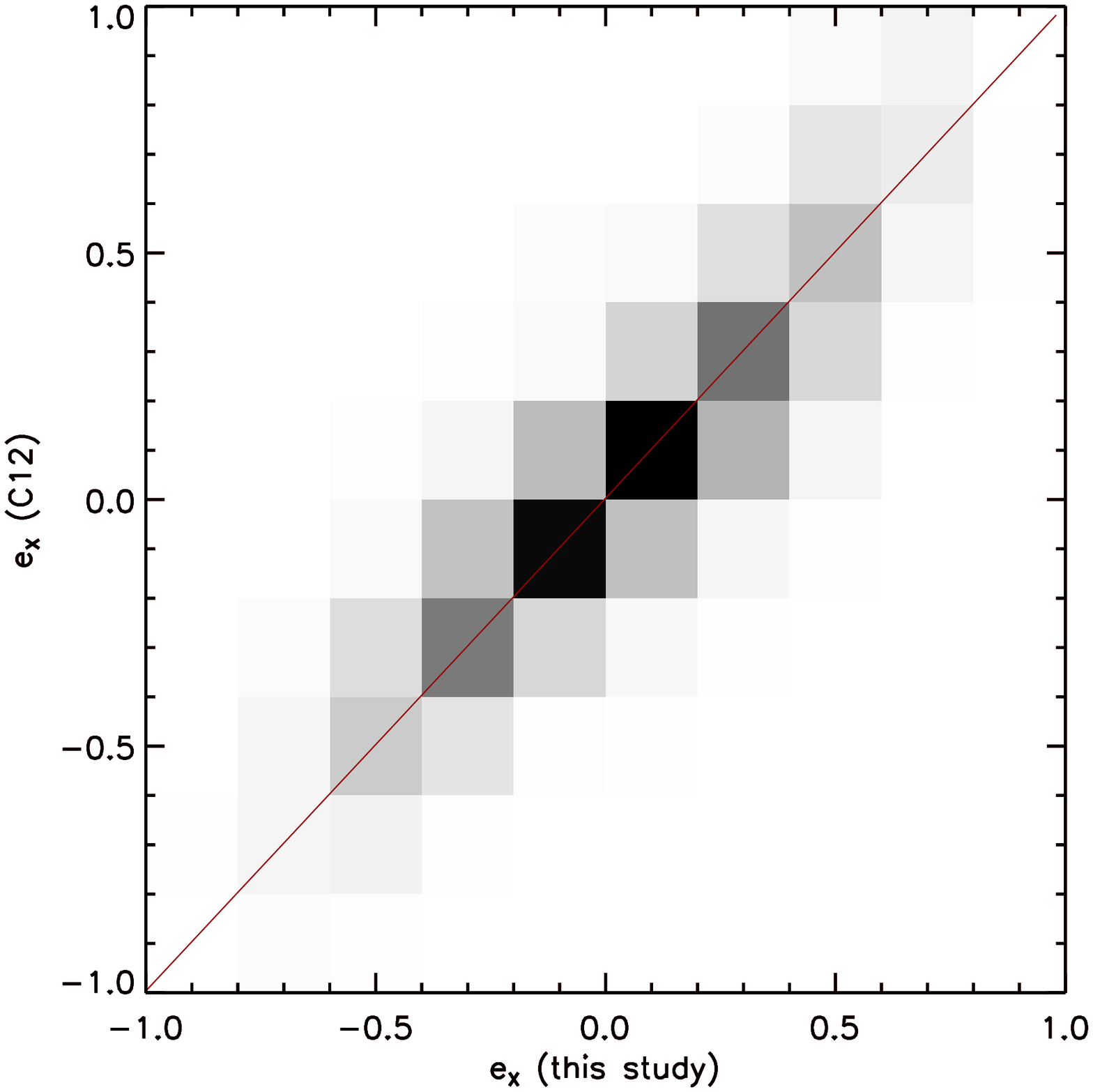}
\end{center}
\caption{Shear measurement comparison between this study and C12.
Shear calibration is applied. A total of 2,148 sources are common to
both shape catalogs. The comparison provides basic sanity checks 
(e.g., systematic difference in shear calibration).
The red solid line is a fit to the data, and we found that the
 mean slope is consistent with unity.
\label{fig_e_com_with_C12}}
\end{figure*}

One of the most basic sanity checks is to compare ellipticities of 
identical sources, and we display the results in Figure~\ref{fig_e_com_with_C12}.
Because the shape catalog provided by C12 already includes their shear calibration, we
also apply our independent (determined from image simulation) shear calibration to our source ellipticity to enable a
fair comparison.  Figure~\ref{fig_e_com_with_C12} shows that there is no major shear calibration difference. The mean
slope is consistent with unity.

\begin{figure*}
\begin{center}
\includegraphics[width=\hsize]{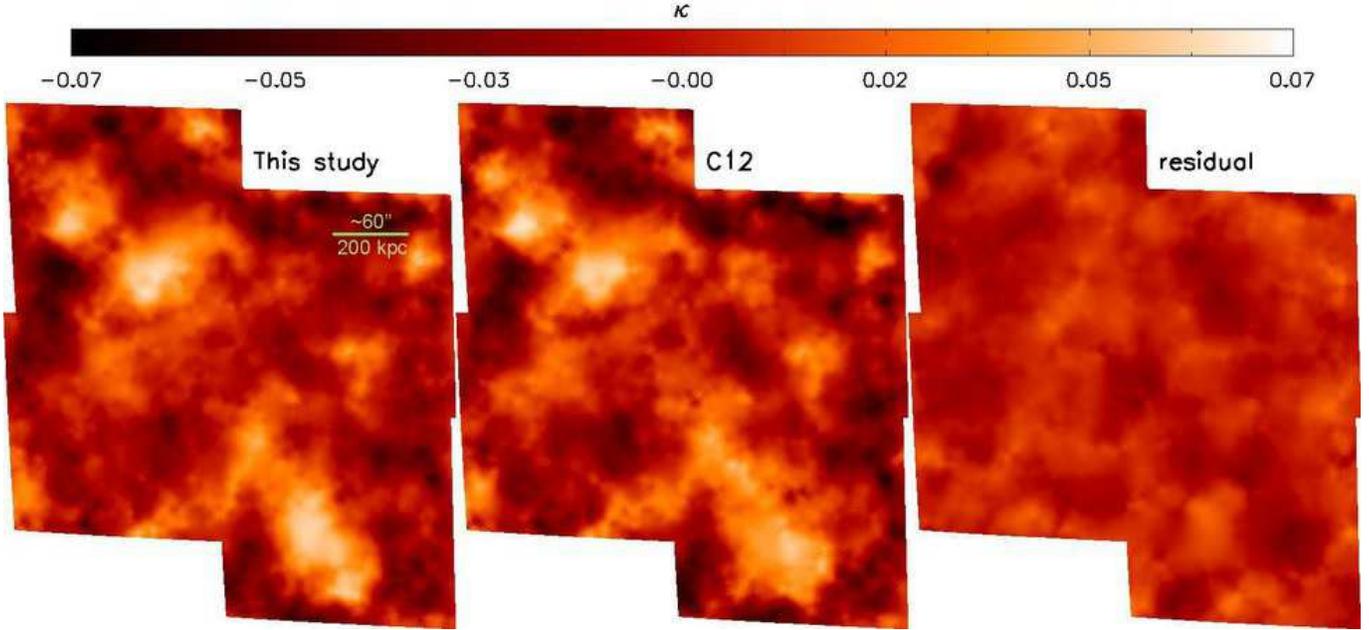}
\end{center}
\caption{Direct comparison of our mass reconstruction with the C12 result.
The mass reconstruction is performed with {\tt FIATMAP} using the 
common 2,148 sources. Note that this number is more than a factor of two smaller
than the total number of sources 4,788 in our study.
The mass overdensity at P3$^{\prime}$ is seen in both mass maps.
\label{fig_map_com_with_C12}}
\end{figure*}

Having found no major systematic difference at least in global shear calibration between the two studies, we compare mass reconstruction results
obtained from the common 2,148 sources. We use the {\tt FIATMAP} code without the nonlinear updating $g=\gamma/(1-\kappa)$ because this may amplify the difference and hamper a fair assessment of the difference.
Figure~\ref{fig_map_com_with_C12} shows the comparison. It is clear that our mass map created with the C12 shape (middle) shows the mass overdensity at P3$^{\prime}$. This result is slightly different from Figure 2 of C12, where the authors perform the mass reconstruction using the combined shapes from HST and Magellan.  Their HST only mass reconstruction is presented in the right panel of Figure 6 in C12, which shows more mass in the dark peak region and hence is more similar to our mass map created with the C12 shapes (middle panel of Figure~\ref{fig_map_com_with_C12}). Our bootstrap experiment with the C12 weak-lensing catalog
shows that a significance mass is found in the dark peak region for their weak-lensing data (Figure~\ref{fig_sig_darkpeak}).
We conclude that the C12 mass map supports the presence of the significant mass in the dark peak region in A520, although the slightly weaker significance might make the overdensity appear as an extension of P4 rather than a separate peak.

\begin{figure}
\begin{center}
\includegraphics[width=\hsize]{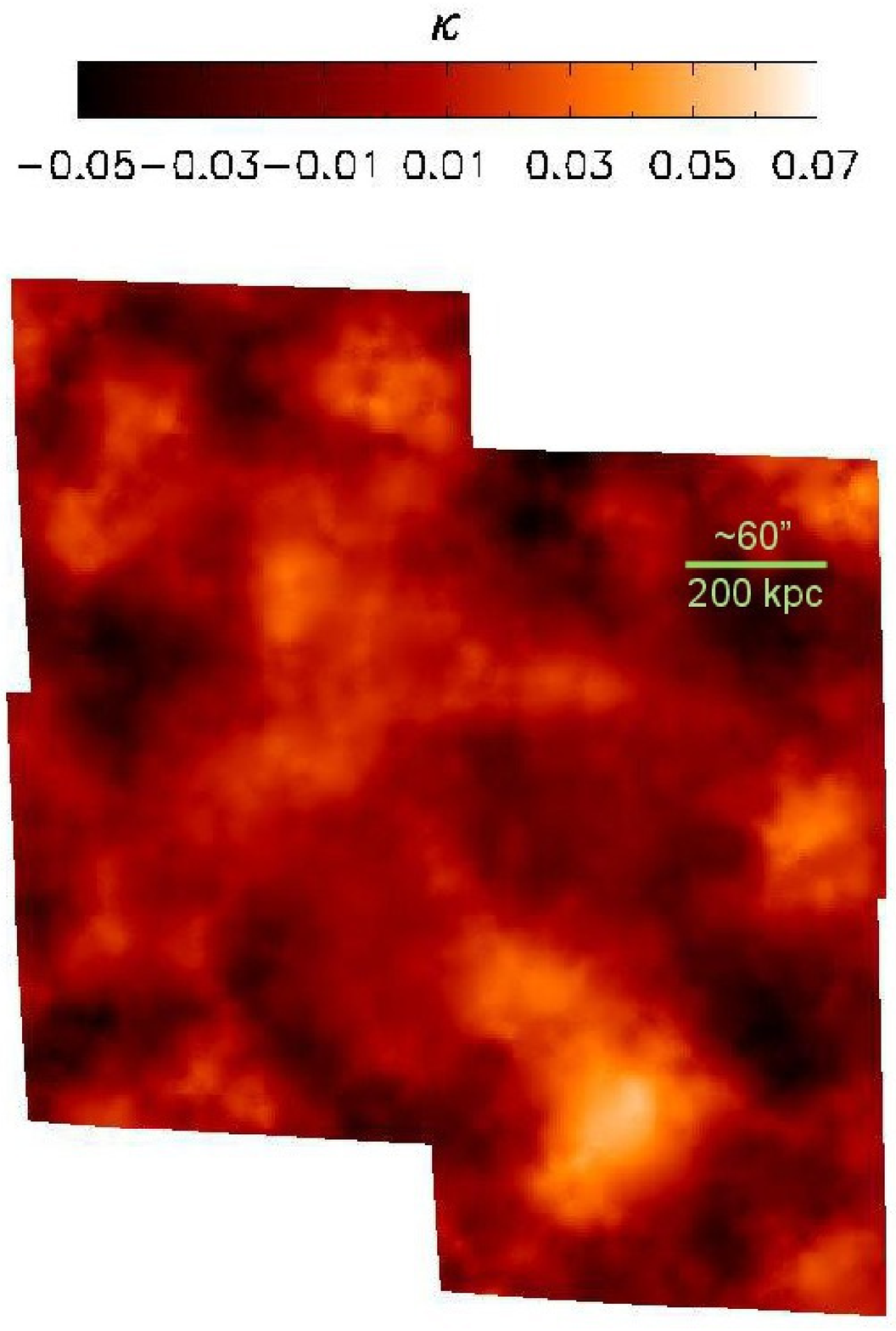}
\end{center}
\caption{Mass reconstructing using the faint galaxies present in the catalog of this study but excluded by C12. We use {\tt FIATMAP} with the remaining 2640 sources.
The resulting mass map is similar to our full mass reconstruction obtained from all 4788 sources. This illustrates that the faint galaxies discarded by C12 but included in the current
study contains significant lensing signals. In addition, the results verifies that the dark peak in A520 is not caused by 
potential systematic shape errors in the low S/N galaxies.
\label{fig_mass_only_in_J13}}
\end{figure}

Remember that the above comparison is limited to the common sources found in both the current and C12 studies. Thus, an important question is how much the source galaxies not used by C12 (but included in our study) affect the results. In Figure~\ref{fig_mass_only_in_J13}, we present the mass reconstruction from these sources. Note that this is  a completely independent mass map.
The result is similar to the one presented in Figure~\ref{fig_map_com_with_C12}. We can see the same substructures (including the dark peak) in this version. This illustrates that the faint galaxies discarded by C12 but included in the current
study contains significant lensing signals. In addition, the comparison verifies that the dark peak in A520 is not caused by any
potential systematic shape errors in the low S/N galaxies.

\section{Interpretation of the Dark Peak} \label{section_interpretation}

Our ACS weak lensing analysis supports the finding of M07 and J12 of a
significant amount of dark matter in A520 at a location where there
are few luminous cluster galaxies. As in J12 we identify a peak, but
its position has shifted (as explained in \S4.3) and now coincides well
with the peak in the X-ray emission. The M/L value within the r=150
kpc aperture of this new centroid is estimated to be $813\pm78
M_{\sun}/L_{R_{\sun}}$ after the gas mass is subtracted. This value is
much larger than what is typically observed in clusters and either points
to a pile-up of dark matter or a reduction in cluster galaxies in that
region. M07 and J12 discussed a number of scenarios, and readers are
referred to these two papers for details.

Here we revisit the subject of collisional dark matter as a potential
origin of this dark substructure. Our motivation is not that the
current ACS analysis result favors this scenario, but that it is worth
investigating it given the updated centroid and substructure
properties. We do note that the new location of the dark peak is in
better agreement with the densest part of the X-ray emission and the
morphology of this substructure is extended along the merger axis
toward P2, which may support the collisional dark-matter hypothesis
for the nature of the substructure.

M07 assumed a toy model, where peaks 1, 2, 4, and 5 each contributed
$\mytilde25$\% of the total mass observed in the central dark peak.
Thus, the model assumes that the chance of dark matter scatter per
particle during this encounter is 1 in 4 or

\begin{equation}
\tau = \frac{\sigma_{DM}}{m_{DM}} \Sigma_M \approx 0.25, \label{eqn_dm_crosssection}
\end{equation}
\noindent where $\Sigma_M$ is the effective scattering depth viewed by
a particle moving along the merger axis.  M07 used the mass of P3 to
estimate this effective depth, and obtained $\sigma_{DM}/m_{DM} \sim
3.8 \pm1.1~\mbox{cm}^2 \mbox{g}^{-1} $, which is about 4 $\sigma$
higher than the $\sigma_{DM}/m_{DM} < 1~\mbox{cm}^2 \mbox{g}^{-1} $
constraint from the Bullet Cluster (Markevitch et al. 2003; Randall et
al. 2008).

In this paper, we revise the M07 model as follows. We decompose the
mass of the dark peak into the contributions from the P2 and P4 halos,
the gas mass, the dark matter associated with the cluster galaxies
found within the $r=150$kpc aperture, and the excess dark matter due
to self-interaction. On the other hand, M07 considered the possibility
that the entire mass of the dark peak originates from
self-interaction.

Our ACS mass reconstruction indicates that the overall pre-merger
cluster mass distribution might be approximately bimodal, dominated by
two massive halos (P2 and P4) when we hypothesize that the dark peak
in the center is produced after the collision. We estimate the
contribution from the wings of these two halos by assuming an NFW
profile with a scale radius of 100 kpc. The second parameters of the
NFW model are determined by the lensing masses within the $r=150$ kpc
aperture centered on each halo.  The total contribution to the
P3$^{\prime}$ mass is determined to be $\sim1.4\times10^{13}M_{\sun}$
with P2 and P4 providing $\sim0.2\times10^{13}M_{\sun}$ and
$\sim1.2\times10^{13}M_{\sun}$, respectively. Using the
Cauchy-Schwarts inequality, we obtain a generous upper limit of
$0.85\times10^{13}M_{\sun}$ for the plasma mass within the $r=150$kpc
radius of P3$^{\prime}$.  The $R$-band luminosity of
$0.38\times10^{11} L_{R\sun}$ is converted to
$0.43\times10^{13}M_{\sun}$ using the average M/L of the rest of the
subclusters, and we assume this to represent the dark matter mass
associated with the few cluster galaxies around the region. Then, the
net excess mass attributed to the collisional deposit becomes
$\sim1.26\times10^{13}M_{\sun}$.

Now the most uncertain part of this scenario is how this excess dark
matter is contributed by the subclusters of A520, and this dominates
our uncertainty in the estimation of the collisional cross-section.
If we assume P2 and P4 contribute equally to this excess mass through
dark matter self-interaction, the mass loss fraction for each substructure
is $\sim13$\%.  The lower limit of the scattering depth $\Sigma_M$ is
the surface mass density of P2 or P4 before the mass loss (we assume
spherical symmetry). In this case, we obtain $\Sigma_{M}=0.138\pm0.009
\mbox{g}\mbox{ cm}^{-2}$ by averaging the surface mass density of P2 and P4
and multiplying the result by 1.13.  Then, from the scattering
probability of $\sim13$\% we estimate $\sigma_{DM}/m_{DM}\approx 0.13/
(0.138\pm0.009) \mbox{cm}^{2} \mbox{g}^{-1} \approx 0.94\pm0.06 \mbox{
  cm}^{2} \mbox{g}^{-1}$.  This value does not violate the Bullet
Cluster estimate $\sigma_{DM}/m_{DM} \leq 1 \mbox{ cm}^{2}
\mbox{g}^{-1}$ of Markevitch et al. (2003).  Of course, the required
cross-section decreases if we assume that the scattering depth is
higher than the adopted value.  This would happen if the dark matter
particles in P4 have passed through more than P2 before it arrived at
the current observed location.  In fact, the location of the bow-shock
feature indicates that P4 may correspond to the ``bullet'' of the
Bullet Cluster and have experienced the most acceleration, which implies that the
other gravitational potential may have been deeper than that of P2. If
we assume that P4 passed through P1, P2, and P5 on its way to the
current location, the scattering depth increases to
$\Sigma_{M}=0.302\pm0.019 \mbox{g}\mbox{ cm}^{-2}$, which gives a
cross-section $\sigma_{DM}/m_{DM}\approx 0.43\pm0.12 \mbox{ cm}^{2}
\mbox{g}^{-1}$.

Although the above constraint on the dark matter cross-section is
based on simplistic assumptions of the unobserved pre-merger
configuration, the result is interesting because it shows that the
current observation of A520 may be explained by self-interaction of
dark matter without creating any serious tension with previous
values. Tighter constraints may become possible if the cluster is
followed up with detailed numerical studies.

\section{CONCLUSIONS}  \label{section_conclusion}

We have presented a re-analysis of HST/ACS images of A520.  Our ACS
weak lensing study confirm the presence of a region of very high
mass-to-light ratio, first reported in M07 with CFHT
data and subsequently supported by Obake \& Umetsu (2008) and J12
with Subaru and WFPC2 data, respectively. We are able to
reproduce the results from J12 when we match the selection of our ACS
weak lensing catalog to that of the WFPC2 analysis, but find no clear
peak at the location where they reported one. Our detailed comparison
suggest that this is caused by a variation in the source number
density, which leads to additional systematic noise in the mass map. The ACS
analysis shows less variation, owing in part to the overall higher
number density, and thus should be more reliable.
 
The analysis presented here indicates a peak that is shifted by $\sim
1'$ compared to J12. Its position now coincides well with the location
of the peak of the X-ray emission. Our mass reconstruction compares
well with that of C12, although we identify a number
of differences. In particular, C12 do not identify such a clear peak,
although we note an extension of P4 in their map towards P3$^\prime$.
A comparison of CTI correction algorithms, including one used by C12,
suggest that the density contrast at the location of P3$^\prime$ is
affected by CTI (see Figure~\ref{fig_cti_mass_reconstruction}).  We
use the latest algorithm from Ubeda \& Anderson (2012) which performs
best as demonstrated in Figure~\ref{fig_cti_ellipticity}. Note that 
this CTI correction method was not available to C12.

Our shape measurement analysis is able to reach a source number
density of $\mytilde109$ arcmin$^{-2}$, which is considerably higher
than the $\mytilde56$ arcmin$^{-2}$ used by C12. This may be explained
by differences in the reduction of ACS data and how measurements in
the different filters are combined. The three-filter ACS data allow
for an improved membership determination which increases the
luminosity by $\sim 16\%$ compared to J12. We find that our luminosity
estimates are consistent with C12 when we compare to the same band.
The mass-to-light ratios (after subtracting the X-ray gas mass) of the
dark peak using the old and new centroids are
$285\pm34M_{\sun}/L_{R\sun}$ and $813\pm78M_{\sun}/L_{R\sun}$,
respectively (in the rest-frame $B$-band, $349\pm43M_{\sun}/L_{B\sun}$
and $966\pm97M_{\sun}/L_{B\sun}$, respectively). Our $\chi^2$ test
shows that the constant mass-to-light ratio hypothesis is rejected at
least at the $\mytilde6~\sigma$ level. 
The mass-to-light ratio is therefore much higher than is typically
observed in clusters and could be due to a reduction in cluster
galaxies or an increase in the amount of dark matter in that
region. Although we still cannot single out a scenario that explains
the observations, we revisit the case of collisional dark matter. With
the updated substructure properties and consideration of other
physical factors for the contribution to the dark peak mass, we find
that the net excess mass of the dark peak region can be explained with
a more conventional range of dark matter self-interacting cross-section
$\sigma_{DM}/m_{DM} \approx 0.43-0.94 \mbox{ cm}^{2} \mbox{g}^{-1}$,
where the uncertainty is dominated by unknown scattering depth along
the merger axis.  This range is consistent with the results obtained from the Bullet Cluster.
Detailed numerical simulations must be carried out
to draw more physically meaningful constraints from the current A520
observation.  Nevertheless, our analytic study hints at the
possibility that A520 can be used to investigate the lower limit of
the self-interacting dark matter.

\acknowledgments
We thank D. Clowe for some useful discussions and agreeing to exchange weak-lensing catalogs.
M.~J. acknowledges support from the National Science Foundation under Grant No. PHYS-1066293 and the hospitality of the Aspen Center for Physics.
H.~H. ackowledges support from NWO VIDI grant number 639.042.814.
This research was supported in part by the National Science Foundation under Grant No. NSF PHY05-51164 and NSF PHY11-25915 to KITP.   AB acknowledge support from NSERC Canada through the Discovery Grant program.

\end{document}